\documentclass[superscriptaddress,preprint,secnumarabic,amssymb,nobibnotes,nofootinbib,aps,prc]{revtex4}
\usepackage{graphicx}
\usepackage{bm}
\newcommand{\bea}{\begin{eqnarray}}
\newcommand{\eea}{\end{eqnarray}}
\newcommand{\be}{\begin{equation}}
\newcommand{\ee}{\end{equation}}
\newcommand{\bt}{\begin{tabular}}
\newcommand{\et}{\end{tabular}}
\newcommand{\Tr}{{\rm Tr}}
\newcommand{\no}{\nonumber}
\newcommand{\ovl}{\overline}

\newcommand{\pa}{\partial}
\newcommand{\beas}{\begin{eqnarray*}}
\newcommand{\eeas}{\end{eqnarray*}}
\begin{document}
\date{\today}
\title{In-medium vector meson masses in a Chiral SU(3) model}
\author{D. Zschiesche}
\email{ziesche@th.physik.uni-frankfurt.de}
\affiliation{Institut f\"ur Theoretische Physik,
        Robert Mayer Str. 8-10, D-60054 Frankfurt am Main, Germany}

\author{A.Mishra}
\email{mishra@th.physik.uni-frankfurt.de}
\affiliation{Institut f\"ur Theoretische Physik,
        Robert Mayer Str. 8-10, D-60054 Frankfurt am Main, Germany}

\author{S. Schramm} 
\affiliation {Argonne National Laboratory, 
9700 S. Cass Avenue, Argonne IL 60439, USA}


\author{H.~St\"ocker}
\affiliation{Institut f\"ur Theoretische Physik,
        Robert Mayer Str. 8-10, D-60054 Frankfurt am Main, Germany}

\author{W.~Greiner}
\affiliation{Institut f\"ur Theoretische Physik,
        Robert Mayer Str. 8-10, D-60054 Frankfurt am Main, Germany}

\begin{abstract}
A significant drop of the vector meson masses in nuclear matter
is observed in a chiral SU(3) model due to the effects of the 
baryon Dirac sea. This is taken into account through the summation 
of baryonic tadpole diagrams in the relativistic Hartree approximation. 
The appreciable decrease of the in-medium vector meson masses is due
to the vacuum polarisation effects from the nucleon sector 
and is not observed in the mean field approximation.
 
\end{abstract}
\pacs{}
\maketitle
\def\bfm#1{\mbox{\boldmath $#1$}}

\section{Introduction}
The medium modifications of the vector mesons ($\rho$ and $\omega$)
in hot and dense matter have recently been a topic of great interest
in the strong interaction physics research, both experimentally 
\cite {helios,ceres,dls,rhic,hades}
and theoretically \cite{brown,rapp,hat,jin,samir,weise,ernst}.
One of the explanations of the experimental observation of
enhanced dilepton production
\cite{helios,ceres,dls} in the low invariant mass regime could
be a reduction in the vector meson masses in the medium.
It was first suggested by Brown and Rho that the vector meson masses
drop in the medium according to a simple (BR) scaling law \cite{brown}, 
given as $m_V^*/m_V=f_\pi^*/f_\pi$. $f_\pi$ is the pion decay constant
and the asterisk refers to in-medium quantities.
There have also been QCD sum rule approaches extensively used in the
literature \cite{hat,jin,samir,weise} for consideration of the in-medium
vector meson properties. In the framework of Quantum Hadrodynamics (QHD) 
\cite{qhd} as a description of the hadronic matter, it is seen that
the dropping of the vector meson masses has its dominant
contribution arising from the vacuum polarisation effects in the
baryon sector \cite{hatsuda,hatsuda1,jeans,sourav}.
This drop is not observed in the mean field approximation.
The vector meson properties \cite{vecmass} and their effects
on the low mass dilepton spectra \cite{dlp} have been investigated 
recently including the quantum correction effects from the baryon 
as well as the scalar meson sectors in the Walecka model \cite{mishra}.

In the present investigation we use the SU(3) chiral model 
\cite{paper3,springer} for the description of the hadronic matter.
This model has been shown to successfully describe hadronic
properties in the vacuum as well as nuclear matter,
finite nuclei and neutron star properties.
Furthermore the model consistently includes the lowest
lying baryon and meson multiplets, including the vector mesons. 
In the mean field approximation the vector meson masses do not show
any significant drop, similar to results in
the Walecka model. 
The effect of the Dirac sea is taken into account
by summing over baryonic tadpole diagrams
in the relativistic Hartree approximation (RHA). 
It is seen that an appreciable decrease of the vector meson
masses arises from the nucleon Dirac sea. This shows the importance of
taking into account these contributions. 

We organize the paper as follows: In section \ref{model} 
we introduce the chiral $SU(3)$ model used in the present
investigation. Section \ref{mfa} describes the mean field
approximation for nuclear matter. In section \ref{RHA} the
nuclear matter properties are considered in the relativistic
Hartree approximation.  Section \ref{vmeson} gives the in-medium
vector meson properties due to the contributions from the nucleon Dirac sea. 
The results are presented and discussed in section \ref{results}.
Finally, in section \ref{sum} we summarize the findings of the present
work.

\section{The hadronic chiral $SU(3) \times SU(3)$ model}
\label{model}
We consider a relativistic field theoretical model of baryons and
mesons built on chiral symmetry and broken scale invariance
\cite{paper3,springer}. A nonlinear realization of chiral symmetry
is adopted, that has been successful in a simultaneous description
of finite nuclei and hyperon potentials \cite{paper3}.
The general form of the Lagrangian is as follows:
\be
\label{lagrange}
{\cal L} = {\cal L}_{\mathrm{kin}}+
\sum_{W=X,Y,V,{\cal A},u}{\cal L}_{\mathrm{BW}}+
{\cal L}_{\mathrm{VP}}
+{\cal L}_{\mathrm{vec}}+{\cal L}_{0}+{\cal L}_{\mathrm{SB}} .\no
\ee
${\cal L}_{\mathrm{kin}}$ is
the kinetic energy term, ${\cal L}_{\mathrm{BW}}$ includes the
interaction terms of the baryons with the spin-0 and spin-1
mesons, the former generating the baryon masses.
${\cal L}_{\rm{VP}}$ contains the interaction terms
of vector mesons with pseudoscalar mesons.
${\cal L}_{\rm{vec}}$ generates the masses of the spin-1 mesons through
interactions with spin-0 fields and contains quartic self-interactions
of the vector-fields. ${\cal L}_{0}$ gives the meson-meson
interaction terms which induce the spontaneous breaking of chiral symmetry.
It also includes a scale-invariance breaking logarithmic potential. Finally,
${\cal L}_{\mathrm{SB}}$ introduces an explicit symmetry breaking of the
U(1)$_A$, SU(3)$_V$ and chiral symmetry.

\subsection{Kinetic Terms}
The kinetic energy terms are given as \cite{paper3}
\bea
\label{kinetic}
{\cal L}_{kin} &=& i\Tr \overline{B} \gamma_{\mu} D^{\mu}B
                + \frac{1}{2} \Tr D_{\mu} X D^{\mu} X
+  \Tr (u_{\mu} X u^{\mu}X +X u_{\mu} u^{\mu} X)
                + \frac{1}{2}\Tr D_{\mu} Y D^{\mu} Y \nonumber \\
               &+&\frac {1}{2} D_{\mu} \chi D^{\mu} \chi
                - \frac{ 1 }{ 4 } \Tr
\left(\tilde V_{ \mu \nu } \tilde V^{\mu \nu }  \right)
- \frac{ 1 }{ 4 } \Tr \left(F_{ \mu \nu } F^{\mu \nu }  \right)
- \frac{ 1 }{ 4 } \Tr \left( {\cal A}_{ \mu \nu } {\cal A}^{\mu \nu }  \right),
\eea
where $B$ is the baryon octet, $X$ is the scalar multiplet, $Y$ is the
pseudoscalar chiral singlet, $\tilde V^\mu$ is the vector meson multiplet
with field tensor
$\tilde V_{\mu \nu}=\partial^\nu \tilde V^\mu-\partial^\nu \tilde V^\mu $,
\footnote{As described in section \ref{meson-meson}, 
the vector mesons need to be renormalized. The physical fields 
will be denoted as $V_{\mu}$ and
$\rho_\mu, \omega_\mu, \phi_\mu$ respectively and 
the unrenormalized, mathematical fields as
$\tilde V_{\mu}$ and $\tilde \rho_\mu, \tilde \omega_\mu, \tilde \phi$.}
$A_{\mu \nu}=\partial^\nu A^\mu-\partial^\nu A^\mu $
is the axialvector field tensor, $F_{\mu \nu}$ is the electromagnetic
field tensor and $\chi$ is the scalar, iso-scalar glueball-field.
The kinetic energy term for the pseudoscalar mesons
is given in terms of the axial vector 
$u_\mu=-\frac i2 \left[u^\dagger \partial_\mu u- u \partial_\mu
u^\dagger \right]$, 
where 
$u=\exp \Big [{ \frac {i}{2\sigma_0} \pi^a \lambda_a \gamma_5}\Big ]$
is the unitary transformation operator 
\cite{paper3}. 
The pseudoscalar mesons are given as parameters of the symmetry
transformation. Since the fields in the nonlinear realization of
chiral symmetry contain the local unitary transformation operator,
 covariant derivatives
$D_\mu = \partial_\mu  + i\left[\Gamma_\mu ,\; \right],$ with
$ \Gamma_\mu = -\frac i2 \left[u^\dagger \partial_\mu u + u\partial_\mu u^\dagger \right]$
have to be used to guarantee
chiral invariance \cite{paper3}. E.g. for the baryons this yields
\be
D_\mu B = \partial_\mu B + i\left[\Gamma_\mu ,B\right].
\ee

\subsection{Baryon-Meson interaction}

The $SU(3)$ structure of the
the baryon-meson interaction terms are the same for all mesons,
except for the difference in Lorentz space.
For a general meson field $W$ they read
\be
{\cal L}_{BW} =
-\sqrt{2}g_8^W \left(\alpha_W[\ovl{B}{\cal O}BW]_F+ (1-\alpha_W)
[\ovl{B} {\cal O}B W]_D \right)
- g_1^W \frac{1}{\sqrt{3}} \Tr(\ovl{B}{\cal O} B)\Tr W  \, ,
\ee
with $[\ovl{B}{\cal O}BW]_F:=\Tr(\ovl{B}{\cal O}WB-\ovl{B}{\cal O}BW)$ and
$[\ovl{B}{\cal O}BW]_D:= \Tr(\ovl{B}{\cal O}WB+\ovl{B}{\cal O}BW) - \frac{2}{3}\Tr (\ovl{B}{\cal O} B) \Tr W$.
The different terms to be considered are those for the interaction
of baryons  with
scalar mesons ($W=X, {\cal O}=1$), with
vector mesons  ($W=\tilde V_{\mu}, {\cal O}=\gamma_{\mu}$ for the vector and
$W=\tilde V_{\mu \nu}, {\cal O}=\sigma^{\mu \nu}$ for the tensor
interaction),
with axial vector mesons ($W={\cal A}_\mu, {\cal O}=\gamma_\mu \gamma_5$)
and with
pseudoscalar mesons ($W=u_{\mu},{\cal O}=\gamma_{\mu}\gamma_5$), respectively.
In the following we discuss the relevant couplings
for the current investigation.

\subsubsection{Baryon - scalar meson interaction (Baryon Masses)}
\label{baryonmasses}
The baryons and the scalar mesons transform equally in the left and
right subspaces. Therefore, in contrast to the linear realization of
chiral symmetry, an $f$-type coupling is allowed for the baryon meson
interaction. In addition, it is possible to construct mass terms for
baryons and to couple them to chiral singlets.
After insertion of the vacuum expectation value for the scalar multiplet matrix
$\langle X\rangle_0$, one obtains the baryon masses as
generated by the VEV of the non-strange 
$\sigma \sim \langle \bar u u + \bar d d \rangle$ 
and the strange $\zeta \sim \langle \bar s s \rangle$ scalar fields
 \cite{paper3}.
Here we will consider the limit $\alpha_S=1$ and
$g_1^S=\sqrt{6}g_8^S$.
In this case the nucleon mass
does depend only on the nonstrange condensate $\sigma$. Furthermore,
the coupling constants between the baryons and the two scalar condensates
are related to the additive quark model. This leaves only one coupling
constant free that is adjusted to give the correct nucleon mass
\cite{paper3}. For a fine-tuning of the remaining masses, 
it is necessary to introduce an explicit symmetry breaking term, 
which breaks the SU(3)-symmetry along the hypercharge
direction (for details see \cite{paper3}).
Therefore the resulting baryon octet masses for the current
investigation read:
\bea
\label{bmver2}
m_N &=& -g_{N \sigma}  \sigma_0  \nonumber  \\
m_{\Lambda} &=& -g_{N \sigma}
\left(\frac{2}{3}\sigma_0-\frac{1}{3}\sqrt{2}\zeta_0 \right)
+\frac{m_1+2 m_2}{3} \nonumber \\
m_{\Sigma} &=& -g_{N \sigma}
\left(\frac{2}{3}\sigma_0 -\frac{1}{3}\sqrt{2}\zeta_0\right) +m_1
\nonumber \\
m_{\Xi} &=&-g_{N \sigma}
\left(\frac{1}{3}\sigma_0 -\frac{2}{3} \sqrt{2}\zeta_0\right)+m_1  +m_2.
\eea
Alternative ways of mass generation have also been considered earlier
\cite {paper3}.

\subsubsection{Baryon - vector meson interaction}
\label{bvmeson}
Two independent interaction
terms of baryons with spin-1 mesons can be constructed
in analogy with the baryon-spin-0-meson interaction. They correspond to
the antisymmetric ($f$-type) and symmetric ($d$-type) couplings,
respectively. The general couplings are shown in \cite{paper3}.
From the universality
principle \cite{saku69} and the vector meson dominance model one may
conclude that the $d$-type
coupling should be small. Here we will use pure $f$-type coupling,
i.e. $\alpha_V=1$ for all fits, even though
a small admixture of d-type coupling
allows for some fine-tuning of the single particle energy levels of nucleons
in nuclei (see \cite{paper3}).
As for the case with scalar mesons, we furthermore set
$g_1^V=\sqrt{6}g_8^V$, so that
the strange vector field $\tilde \phi_{\mu} \sim
\ovl{s}\gamma_{\mu} s $
does not couple to the nucleon.
The resulting Lagrangian reads:
\be
\label{baryon-vector-eq}
{\cal L}_{BV} =
-\sqrt{2}g_8^V
\left([\ovl{B}{\cal \gamma_\mu }B \tilde V^\mu]_F +
\Tr(\ovl{B}{\cal \gamma_\mu } B)\Tr \tilde V^\mu \right)  \, ,
\ee
or explicitly written out for the nuclear matter case:
\be
{\cal L}_{BV}^{N} = 3 g_8^V \tilde \omega_\mu
 \overline{\psi_{N}} \gamma_\mu \psi_{N} +
 g_8^V \tilde{\vec{\rho}_\mu}
 \overline{\psi_{N}} \gamma_\mu \vec \tau \psi_{N}.
\ee
Note that in this limit all coupling
constants are fixed once $g_8^V$ is
specified \cite{paper3}.
This is done by fitting the nucleon-$\omega$ coupling  
to the energy density at nuclear
matter saturation ($E/A = -16$~MeV).
Since we consider nuclear matter, the couplings of the vector
mesons to the hyperons shall not be discussed here.

\subsection{Meson-meson interactions}
\label{meson-meson}
\subsubsection{Vector mesons}

The vector meson-meson interactions contain the mass terms of the
vector mesons and higher order vector meson self-interactions.
The simplest scale invariant mass term is
\be
\label{vecfree}
{\cal L}_{vec}^{(1)}= \frac{1}{2} m_V^2 \frac{\chi^2}{\chi_0^2}
\Tr \tilde V_{\mu} \tilde V^{\mu}.
\ee
It implies a mass degeneracy for the vector meson nonet. The scale
invariance is assured by the square of the
glueball field $\chi$  (see sec. \ref{spin0pot} for details).  To split the masses, one can add the chiral invariants 
\cite{gasi69,mitt68}
\be
\label{lvecren}
{\cal L}_{vec}^{(2)} = \frac{1}{4} \mu \Tr
\left[\tilde V_{\mu \nu} \tilde V^{\mu \nu} X ^2 \right]
\ee
and
\be
\label{lvecren2}
{\cal L}_{vec}^{(3)} = \frac{1}{12} \lambda_V
\left(\Tr\left[\tilde V_{\mu \nu} \right]\right)^2.
\ee
Note that in \ref{lvecren}, we replace the scalar multiplet $X$ 
by its vacuum expectation value.
Combining the contributions (\ref{lvecren},\ref{lvecren2})
with the kinetic energy term
(\ref{kinetic}), one obtains the following terms for the vector
mesons in the vacuum
\bea
\label{kinren}
-\frac 14 Z_\rho^{-1} \left( \tilde V_{\tilde \rho}^{\mu\nu} \right)^2
-\frac 14 Z_\omega^{-1} \left( \tilde V_{\tilde \omega}^{\mu\nu} \right)^2
 -\frac 14 Z_\phi^{-1} \left( \tilde V_{\tilde \phi}^{\mu\nu} \right)^2,
\eea
with e.g.
$\tilde V_{\tilde \rho}^{\mu\nu}=\partial^\mu \tilde
\rho^\nu-\partial^\nu \tilde \rho^\mu $.
With the renormalization constants the new vector meson fields 
are defined as
$\rho = Z_\rho^{-1/2} \tilde \rho$,
$\omega = Z_\omega^{-1/2} \tilde \omega$,
$\phi = Z_\phi^{-1/2} \tilde \phi$.
Explicitly the renormalization constants are given as
\bea
\label{zren}
Z_\rho^{-1} = \left(1-\mu \frac{\sigma_0^2}{2}\right);\;\;\;\;
Z_{\omega,\phi}^{-1} =\Bigg [ \left(1- \frac{\mu(\sigma_0^2
+2\xi_0^2)-2\lambda_V}{4} \right ) \pm \frac {1}{2} D^{1/2}\Bigg]
\eea
where
\be
D=\frac {\mu ^2}{4}(\sigma_0^2-2 \xi_0^2)^2+ \lambda_V^2
-\frac {\lambda_V}{3} \mu (\sigma_0^2-2 \xi_0^2).
\ee
Then the Lagrangian for the new fields in the vacuum reads
\bea
{\cal L}_{vec}^{vac}
=
&=&
-\frac 14 \left( \left( V_\rho^{\mu\nu} \right)^2 \right.
+ \left( V_\omega^{\mu\nu} \right)^2
\left. +\left( V_\phi^{\mu\nu} \right)^2 \right)
+\frac 12 \frac {\chi^2}{\chi_0^2}
\left(m_\rho^2 \rho^2 \right.
\left.+ m_\omega^2 \omega^2 + m_\phi^2 \phi^2\right)
\eea
where
\be
m_\rho^2 = Z_{\rho} m_V^2 \quad , \quad
m_{\omega}^2 = Z_{\omega} m_V^2 \quad , \quad
m_{\phi}^2 = Z_{\phi} m_V^2 
\ee
denote the vector meson masses in the vacuum.
Using $m_V=687.33 MeV$, $\mu \sigma_0^2=0.41$ and $\lambda_V=-0.041$,
the correct $\omega$-, $\rho$- and $\phi$-masses are obtained.
The vector meson self-interactions \cite {toki} read
\be
\label{vecquart}
{\cal L}_{vec}^{(4)}=2 (\tilde g_4)^4\,  \Tr (\tilde V_{\mu} \tilde V^{\mu})^2.
\ee
The coupling of this self-interaction term is also modified by
the redefinition of the fields. The redefined coupling corresponding to
the quartic interaction for the $\omega$ field can be expressed in terms
of the coupling $\tilde g_4$ of the term (\ref{vecquart}).
This term gives a contribution to the vector-meson masses in the
medium, i.e. for finite values of the $\omega$ or
$\rho$-fields. The resulting expressions for the vector meson masses in
the medium (isospin symmetric) are
\bea
{m_{\omega}^\ast}^2 &=& 
m_\omega^2 + 12 g_4^4 {\omega}^2
\\
{m_{\rho}^\ast}^2 &=& 
m_\rho^2 + 12 g_4^4 \frac {Z_{\rho}}{Z_\omega} {\omega}^2
\\
{m_{\phi}^\ast}^2 &=& 
m_\phi^2 + 24 g_4^4 \frac{Z_{\phi}^2}{Z_\omega^2} \phi^2,
\eea
with $g_4 = \sqrt{Z_\omega} \tilde g_4$ as the renormalized coupling.
Since the quartic self-interaction contributes only in the medium, 
the coupling $g_4$ cannot be unambigiously fixed. 
It is fitted, so that the compressibility is 
in the desired region between $200-300 \mbox{ MeV}$ in the mean field 
approximation. Note that the $N-\omega$ as well as the
$N-\rho$ - couplings are also affected by the redefinition of the
fields with the corresponding renormalised coupling constants as
$g_{N\omega} \equiv 3 g_V^8 \sqrt{Z_\omega}$ and
$g_{N\rho} \equiv g_V^8 \sqrt{Z_\rho}$.
\subsubsection{Spin-0 Potential}
\label{spin0pot}
In the nonlinear realization of chiral symmetry the couplings of scalar
mesons $X$ and the pseudoscalar singlet $Y$
with each other are only governed by SU(3)$_V$-symmetry.
In this work we will
use the same form of the potential as in the
linear $\sigma$-model with $U(1)_A$ breaking, as described in \cite{paper3}.
It reads
\bea
\label{cpot}
{\cal L}_0 &= &  -\frac{ 1 }{ 2 } k_0 \chi^2 I_2
     + k_1 (I_2)^2 + k_2 I_4 +2 k_3 \chi I_3,
\eea
with $I_2= \Tr (X+iY)^2$, $I_3=\det (X+iY)$ and $I_4 = \Tr (X+iY)^4$.
Furthermore $\chi$ denotes a scalar color-singlet gluon field. It is
introduced to construct the model to satisfy  
the QCD trace anomaly,  
i.e. the nonvanishing of the trace of the energy-momentum tensor
$\theta_{\mu}^{\mu} = \frac{ \beta_{QCD} }{2 g} {\cal G}_{\mu \nu}^a {\cal
G}^{\mu \nu a} $. Here,
${\cal G}_{\mu \nu}^a $ is the gluon field strength tensor of
QCD.

All the terms in the Lagrangian are
multiplied by appropriate powers of
the glueball-field to obtain a dimension $(\mbox{Mass})^4$ in the
fields. 
Then all coupling
constants are dimensionless and therefore the model is scale
invariant \cite{sche80}. Then, a scale breaking potential
\be
\label{lscale}
  {\cal L}_{\mathrm{scalebreak}}=- \frac{1}{4}\chi^4 \ln \frac{ \chi^4 }{ \chi_0^4}
 +\frac{\delta}{3}\chi^4 \ln \frac{I_3}{\det \langle X \rangle_0} 
\ee
is introduced.
This yields
$\theta_{\mu}^{\mu} = (1-\delta)\chi^4$.
By identifying the $\chi$-field with the gluon condensate and the choice
$\delta=6/33$ for three flavors and three colors with 
$\beta_{QCD}$ as given by the one loop level,
the correct trace
anomaly is obtained. The first term in (\ref{lscale})
corresponds to the contribution of the gluons and the second term
describes the
contribution from the quarks to the trace anomaly.
Finally the term 
\be
{\cal L}_{\chi} = - k_4 \chi^4
\ee
generates a phenomenologically consistent 
finite vacuum expectation value.

The parameters
$k_0$,$k_2$ and $k_4$, are used to ensure an extremum in the vacuum for the
$\sigma$-, $\zeta$- and $\chi$-field equations, respectively.
As for the remaining constants,
$k_3$ is constrained by the $\eta$ and $\eta^\prime$-masses, which 
take the values $m_{\eta} = 520 \mbox{ MeV}$ and 
$m_{\eta^\prime} = 999 \mbox{ MeV}$ in all parameter sets.
$k_1$ is fixed in the mean field fit with quartic vector meson
interaction such that the effective nucleon mass at saturation
density is around $0.65 \,m_N$ and the $\sigma$-mass
is of the order of 
500 MeV. Then it is kept constant in
all the other fits, since a change in $k_1$ yields quite a strong
modification of the other coupling constants in the selfconsistency
calculation. Since we want to focus on the
influence of the Hartree terms, we try to keep everything else as
less modified as possible.

Since the shift in the $\chi$ in the medium is rather small
\cite{paper3}, 
we will in good approximation set $\chi=\chi_0$. We will refer to this
case as the {\it frozen glueball limit}.
The VEV of the
gluon condensate, $\chi_0$, is fixed to fit the pressure $p=0$ at the
saturation density $\rho_0 =0.15 \, \rm fm^{-3}$.
\subsection{Explicitly broken chiral symmetry}
In order to eliminate the Goldstone modes from a chiral effective theory,
explicit symmetry breaking terms have to be introduced.
Here, we again take the corresponding term of the
linear $\sigma$-model
\be
\label{esb-gl}
 {\cal L}_{SB}  =
  \frac{1}{2}\Tr A_p(M+M^\dagger)=\Tr A_p\left(u(X+iY)u+u^\dagger(X-iY)u^\dagger\right) 
\ee
with $A_p=1/\sqrt{2}{\mathrm{diag}}(m_{\pi}^2 f_{\pi},m_\pi^2 f_\pi, 2 m_K^2 f_K
-m_{\pi}^2 f_\pi)$ and $m_{\pi}=139$ MeV, $m_K=498$ MeV. This
choice for $A_p$ together with the constraints
\be
\label{zeta0}
\sigma_0 = -f_{\pi} \qquad \zeta_0 = -\frac{1}{\sqrt{2}}(2 f_K - f_{\pi}) ,
\ee
on the VEV on the scalar condensates assure that
the PCAC-relations of the pion and kaon are fulfilled.
With $f_{\pi} = 93.3$~MeV and $f_K = 122$~MeV we obtain $\sigma_0 =
93.3$~MeV and $\zeta_0 = 106.56$~MeV.
\section{Mean Field approximation}
\label{mfa}

The hadronic matter properties at finite density and temperature are studied in
the mean-field approximation \cite{serot97}.
Then the Lagrangian (\ref{lagrange})
becomes
\begin{eqnarray}
{\cal L}_{BX}+{\cal L}_{BV} &=& -\overline{\psi_{N}}[g_{N
\omega}\gamma_0 \omega +m_N^{\ast} ]\psi_{N} \\ 
{\cal L}_{vec} &=& \frac{ 1 }{ 2 } m_{\omega}^{2}\frac{\chi^2}{\chi_0^2}\omega^
2  + g_4^4 \omega^4 \\
{\cal V}_0 &=& \frac{ 1 }{ 2 } k_0 \chi^2 (\sigma^2+\zeta^2)
- k_1 (\sigma^2+\zeta^2)^2
     - k_2 ( \frac{ \sigma^4}{ 2 } + \zeta^4)
     - k_3 \chi \sigma^2 \zeta \nonumber \\
&+& k_4 \chi^4 + \frac{1}{4}\chi^4 \ln \frac{ \chi^4 }{ \chi_0^4}
 -\frac{\delta}{3} \chi^4 \ln \frac{\sigma^2\zeta}{\sigma_0^2 \zeta_0} \\
{\cal V}_{SB} &=& \left(\frac{\chi}{\chi_0}\right)^{2}\left[m_{\pi}^2 f_{\pi}
\sigma
+ (\sqrt{2}m_K^2 f_K - \frac{ 1 }{ \sqrt{2} } m_{\pi}^2 f_{\pi})\zeta
\right],
\end{eqnarray}
where $m_N^*$ is the effective mass of the nucleon.
Only 
the scalar (${\cal L}_{BX}$) and the vector meson terms (${\cal L}_{BV}$)
contribute to the baryon-meson interaction. For all other mesons, 
the expectation value vanishes in the mean-field approximation.
Now it is straightforward to write down the expression
for the thermodynamical potential of the grand canonical
ensemble, $\Omega$, per volume $V$
at a given chemical potential $\mu$ and at zero temperature:
\be
   \frac{\Omega}{V}= -{\cal L}_{vec} - {\cal L}_0 - {\cal L}_{SB}
-{\cal V}_{vac}+ \frac{\gamma_N }{(2 \pi)^3}
\int_0^{\sqrt{{\mu_N^\ast}^2-{m_N^\ast}^2}} d^3k\, [E^{\ast}_N(k)-\mu^{\ast}_N]
\ee
The vacuum energy ${\cal V}_{vac}$ (the potential at $\rho=0$)
has been subtracted in
order to get a energy at $\rho =0$. The factor $\gamma_N$
denotes the fermionic spin-isospin degeneracy factor, and $\gamma_N$
=4 for symmetric nuclear matter. The single particle energy is
$E^{\ast}_N (k) = \sqrt{ k_N^2+{m_N^*}^2}$
and the effective chemical potential reads
 $\mu^{\ast}_N = \mu_N-g_{N\omega} \omega$. \\
The mesonic fields are determined by extremizing the thermodynamic
potential. Since we use the frozen glueball approximation
(i.e $\chi=\chi_0$), we have coupled equations only for the fields
$\sigma$, $\zeta$ and $\omega$ in the selfconsistent calculation 
given as 
\bea
%
\label{sigmft}
\frac{\partial (\Omega/V)}{\partial \sigma} &=&
 k_0 \chi^2 \sigma - 4 k_1 (\sigma^2+\zeta^2)\sigma
 - 2k_2 \sigma^3        - 2 k_3 \chi \sigma \zeta
  -2\frac{\delta \chi^4}{3 \sigma} + \nonumber \\
&+& m_{\pi}^2 f_{\pi}
+ \frac{\pa m_N^{\ast}}{\pa \sigma}\rho^s_N=0, \\
\frac{\partial (\Omega/V)}{\partial \zeta} &=&
   k_0 \chi^2 \zeta - 4 k_1 (\sigma^2+\zeta^2) \zeta
- 4 k_2 \zeta ^3 - k_3 \chi \sigma^2
  -\frac{\delta \chi^4}{3 \zeta} +\nonumber \\ 
 &+& \left[\sqrt{2}m_K^2 f_K
- \frac{ 1 }{ \sqrt{2}} m_{\pi}^2 f_{\pi}\right]=0, \\ 
\frac{\partial (\Omega/V)}{\partial \omega} &=&
-m_{\omega}^2 \omega - 4 g_4^4 \omega^3 + g_{N \omega}\rho_N = 0 .
\eea
In the above, $\rho^s_N$ and $\rho_N$ are the scalar and vector densities
for the nucleons, which can be calculated analytically for the
case of $T=0$, yielding
\bea
\rho^s_N &=& \gamma_N
\int \frac{d^3 k}{(2 \pi)^3} \frac{m_N^{\ast}}{E^{\ast}_N} =
\frac{\gamma_N  m_N^{\ast}}{4 \pi^2}\left[ k_{F N} E_{F N}^{\ast}-m_N^{\ast 2}
\ln\left(\frac{k_{F N}+E_{F N}^{\ast}}{m_N^{\ast}}\right)\right], \\
 \rho_N &=&  \gamma_N \int_0^{k_{F N}} \frac{d^3 k}{(2 \pi)^3} =
\frac{\gamma_N k_{F N}^3}{6 \pi^2}        \,   .
\eea
The parameters of the model are constrained by symmetry relations,
characteristics of the vacuum or nuclear matter properties.
Table \ref{mfparameter} summarizes the various constraints for the parameters within the
Mean-Field approach.
\begin{table}[h]
\begin{center}
\bt{|c|c|c|c|}
\hline
Parameter & Interaction & Lagrange-term & Observable/Constraint\\
\hline
\hline
\begin{tabular}{c}
$g_8^S$
\end{tabular}
&
\begin{tabular}{c}
$\cal{L}_{BM}$
\end{tabular}
&
\begin{tabular}{c}
$g_8^S \sqrt{2} \left(\Tr(\ovl{B} B)\Tr X+ [\ovl{B}BX]_F \right)$
\end{tabular}
&
\begin{tabular}{l}
$m_N = -g_{N \sigma}  \sigma_0$
\end{tabular} \\
\hline
%
%
\begin{tabular}{c}
$g_8^V$ \\
\end{tabular}
&
\begin{tabular}{c}
$\cal{L}_{BV}$  \\
\end{tabular}
&
\begin{tabular}{c}
$g_8^V \sqrt{2}$
$\left(\Tr(\ovl{B} \gamma_\mu B)\Tr V^\mu + [\ovl{B}\gamma_\mu B V^\mu]_F \right) $    \\
\end{tabular}
&
\begin{tabular}{c}
$E/A (\rho_0) = -16 \mbox{ MeV}, \quad  g_{N\omega} \equiv 3 g_8^V Z_\omega$\\
\end{tabular} \\
\hline
%
%
$m_V$ & $ {\cal L}_{vec}^{(1)}$ &
$\frac{1}{2} m_V^2 \frac{\chi^2}{\chi_0^2} \Tr V_{\mu} V^{\mu}  $
& \\
$\mu$ &
$ {\cal L}_{vec}^{(2)}$ &
$\frac{1}{4} \mu \Tr\left[V_{\mu \nu} V^{\mu \nu} X^2 \right]$
&  $m_\omega, m_\rho, m_\phi$\\
$\lambda_V $ & $ {\cal L}_{vec}^{(3)}$ &
$\frac{1}{4} \lambda_V \left(\Tr\left[V_{\mu \nu} \right]\right)^2$ & \\
$g_4$ & $ {\cal L}_{vec}^{(4)}$ & $ 2 g_4^4 \Tr V_{\mu} V^{\mu} $
& $K \approx 200-300 $~MeV \\
\hline
$k_0$ & & $-\frac 12 k_0 \chi^2 I_2$&
$\frac {\partial \Omega}{\partial \sigma} |_{vac} = 0 $\\
$k_1$ & & $k_1 (I_2)^2$ & $m_N^\ast/m_N$, $m_\sigma$\\
$k_2$ & scalar & $k_2 I_4$ &
$\frac {\partial \Omega}{\partial \zeta} |_{vac} = 0 $\\
$k_3$ & potential & $2 k_3 \chi I_3$ & $\eta,\eta^\prime$ masses\\
$k_4$ & & $-k_4 \chi^4$ &$\frac {\partial \Omega}{\partial \chi} |_{vac} = 0 $\\
$\delta$ & & $\frac {\delta}{3} \chi^4 \ln{\frac {I_3}{\det{X}}}$& $\beta_{QCD}$\\
$\chi_0$ & & & $p(\rho_0)$=0 \\
\hline
$m_\pi, m_K$ & & $$ &  \\
$\sigma_0, \zeta_0$ &\raisebox{1.5ex}[-1.5ex] {${\cal L}_{esb}$}
& \raisebox{1.5ex}[-1.5ex]{$-\frac{1}{2} \Tr A_p \left(u(X+iY)u+u^{\dagger}(X+iY)u^{\dagger}\right)$}  &
\raisebox{1.5ex}[-1.5ex] {PCAC} \\
\hline
 \et
\caption{\label{mfparameter} \textrm{
Parameters of the model, the corresponding terms in the Lagrangian and
constraints for fixing them.}}
\end{center}
\end{table}
\section{Relativistic Hartree Approximation}
\label{RHA}
If we go from the Mean-Field to the Hartree
approximation, additional terms in the grand canonical potential
appear. These influence the energy, the pressure and the meson field
equations. In the present work, we use the version of the chiral model with
$g_{N\zeta}=0$, i.e. no coupling of the strange condensate to the
nucleon. Hence additional terms will only appear due to summing over baryonic
tadpole diagrams due to interaction with the scalar field $\sigma$,
similar to as in the Walecka model.
The additional contribution to the energy density is given as
\begin{eqnarray}
\Delta \epsilon &=& -\frac{\gamma _N}{16\pi^2}
\biggl( {m_N^*}^4 \ln \Big (\frac{m_N^*}{m_N}\Big )
 +m_N^3 (m_N-m_N^*)-\frac{7}{2} m_N^2 (m_N-m_N^*)^2 \nonumber\\
 && +  \frac{13}{3} m_N (m_N-m_N^*)^3 -
  \frac{25}{12} (m_N-m_N^*)^4 \biggr),
\end{eqnarray}
where $m_N^*=-g_{\sigma N} \sigma$ and $m_N$ is the nucleon mass in
vacuum. This will also modify the pressure
and the $\sigma$ field equations. With inclusion of the relativistic
Hartree contributions, the field equation for $\sigma$ as given by
(\ref{sigmft}) gets modified to 
\bea
\label{sigrha}
%
 k_0 \chi^2 \sigma - 4 k_1 (\sigma^2+\zeta^2)\sigma
 - 2k_2 \sigma^3        - 2 k_3 \chi \sigma \zeta
  -2\frac{\delta \chi^4}{3 \sigma} + 
m_{\pi}^2 f_{\pi}
+ \frac{\pa m_N^{\ast}}{\pa \sigma}(\rho^s_N+\Delta \rho^s_N)
=0 ,
\eea
where the additional contribution to the nucleon scalar density is
given as
\begin{equation}
\Delta \rho^s_N =
-\frac {\gamma_N}{4\pi^2}
\left [ {m_N^*}^3 \ln \left(\frac{m_N^*}{m_N}\right)
+m_N^2 (m_N-m_N^*)  -  \frac{5}{2} m_N (m_N-m_N^*)^2
 + \frac{11}{6} (m_N-m_N^*)^3 \right]. 
\end{equation}


These make a refitting of some of the parameters
necessary. First we have to account for the change in the energy and
the pressure, i.e. $g_{N\omega}$ and $\chi_0$ have to be refitted. 
Due to a change in $\chi_0$ the parameters $k_0, k_2$ and $k_4$
must be adapted to ensure that the vacuum equations for $\sigma,\zeta$
and $\chi$ have minima at the vacuum expectation values of the fields.
Table \ref{parmfhart} shows the parameters corresponding to the
the Mean-field and the Hartree approximations.
\begin{table}[hc]
\begin{center}
\bt{|c|cc|cc|}
\hline
Parameter & Mean & Field  & Hartree & \\
\hline
$g_4$ & 2.7 & 0 & 2.7 & 0 \\
$k_1$ & 1.4 & 1.4 & 1.4 & 1.4 \\
\hline
$g_{N\omega}$ & 12.83 & 10.52 & 10.51 & 9.37\\
$g_{N\rho}$ & 4.27 & 3.51& 3.50 & 3.12 \\
$\chi_0 $ & 402.7 & 430.1 & 430.2 & 446 \\
$k_3$ & -2.64 & -2.07 & -2.07 & -1.73 \\ 
\hline
$k_0$ & 2.37 & 2.07 & 2.07 & 1.93 \\
$k_2$ & -5.55 & -5.55 & -5.55 & -5.55 \\
$k_4$ & -0.23 & -0.23 & -0.23 & -0.24 \\
$m_N^\ast/m_N (\rho_0)$ &0.64 & 0.71 & 0.73 & 0.76\\
$m_\sigma $ & 475.6 & 560.2 & 560.4 & 610.7 \\
$K$ & 266.1 & 359.5 & 304 & 377.8\\
$a_4$ & 29.0 & 27.4 & 23.9 & 24.1  \\
\hline
\hline
\end{tabular}
\caption{\label{parmfhart} Parameters for the Mean-Field and
the Hartree Fit}
\end{center}
\end{table}

\section{Vector meson properties in the medium}
\label{vmeson}

\subsection {In-medium vector meson masses}

We now examine how the Dirac sea effects discussed in section 
\ref{RHA} modify the masses of the vector mesons.
Rewriting the expression for the vector interaction of these mesons given
in equation (\ref{baryon-vector-eq}) in terms of the renormalized couplings 
$g_{N\omega}$ and $g_{N\rho}$ yields 
\be
{\cal L}_{BV}^{N} = g_{N\omega} \omega_\mu
 \overline{\psi_{N}} \gamma_\mu \psi_{N} +
 g_{N\rho} \vec{\rho}_\mu
 \overline{\psi_{N}} \gamma_\mu \vec{\tau} \psi_{N}.
\ee
Furthermore a tensor coupling is introduced:
\begin{equation}
{\cal L}_{\rm tensor}= - \frac { g_{NV} \kappa_V}{2 m_N} 
\left[\bar \psi_N \sigma_{\mu \nu} \tau^a \psi_N \partial ^\nu V^\mu_a \right]
\label{lint}
\end{equation}
\noindent where $(g_{NV},\kappa_V)=(g_{N\omega},\kappa_\omega)$ 
or $(g_{N\rho},\kappa_\rho)$ and $\tau_a =1$ or $\vec{\tau}$, 
for $V_a^\mu = \omega^\mu$ or $\rho_a^\mu$,
 $\vec{\tau}$ being the Pauli matrices. 
The vector meson self energy is given as
\begin{equation}
\Pi_V^{\mu \nu} (k)=-\gamma_I g_{NV}^2 \frac {i}{(2\pi)^4}\int d^4 p\,
{\rm Tr} \Big [ \Gamma_V^\mu (k) G(p) \Gamma_V^\nu (-k) G(p+k)\Big],
\end{equation}
where $\gamma_I=2$ is the isospin degeneracy factor for nuclear matter, 
and, $\Gamma_V^\mu (k)=
\gamma^\mu\tau_a-({\kappa_{V}}/{ 2 m_N})\sigma^{\mu \nu}\tau_a$ 
represents the meson-nucleon vertex function.
In the above, $G(k)$ is the interacting nucleon propagator
resulting from summing over baryonic tadpole diagrams
in the Hartree approximation.
This is expressed, in terms of the Feynman and density dependent parts,
as
\begin{eqnarray}
G(k) &=& \left (\gamma_\mu \bar{k}^\mu + m_N^\ast\right)
\left[ \frac{1}{\bar{k}^2 - m_N^{\ast 2} + i\epsilon}\right.
\left.+\frac {i\pi}{E_N^\ast(k)} \delta \left({\bar k}^0-E_N^\ast(k)\right) 
\theta\left(k_F - \left| \vec{\bar{k}} \right| \right)\right]
\nonumber \\
&\equiv& G_F(k) + G_D(k).
\end{eqnarray}

The vector meson self energy can then be written as the sum of two parts
\begin{equation}
\Pi^{\mu \nu}= \Pi^{\mu \nu}_F+ \Pi^{\mu \nu}_D.
\end{equation}
In the above, $\Pi^{\mu \nu}_F$ is the contribution arising from the vacuum
fluctuation effects, described by the coupling to the $N\bar N$
excitations and $\Pi^{\mu \nu}_D$ is the density dependent
contribution to the vector self energy. For the $\omega$
meson, the tensor coupling is generally small as compared to the
vector coupling to the nucleons \cite {hatsuda1}. This is 
neglected in the present calculations. The Feynman part of the self energy,
$\Pi ^ {\mu \nu}_F$, is divergent and needs renormalization.
We use dimensional regularization to separate the divergent parts.
For the $\rho$-meson with tensor interactions, 
a phenomenological subtraction procedure \cite {hatsuda,hatsuda1} 
is adopted. 
After renormalisation, the contributions to the meson self energies from the
Feynman part are given as follows. For the $\omega$ meson,
one arrives at the expression
\begin{equation}
\Pi_F ^\omega (k^2) \equiv \frac {1}{3} {\rm Re} (\Pi^{\rm ren}_F)^\mu _\mu
 =-\frac {g_{N\omega} ^2}{\pi^2} k^2
\int _0 ^1 dz z (1-z) \ln \left[ \frac {{m_N^*}^2-k^2 z (1-z)}{{m_N}^2
-k^2 z (1-z)} \right].
\end {equation}
and for the $\rho$ meson,

\begin{equation}
\Pi ^\rho _F (k^2)=-\frac {g_{N\rho}^2}{\pi^2} k^2
\Big [ I_1 +m_N^* \frac {\kappa_\rho}{2 m_N} I_2 +\frac {1}{2}
\Big (\frac {\kappa_ \rho}{2 m_N} \Big )^2 (k^2 I_1+{m_N^*}^2 I_2)
\Big]
\end{equation}
where,
\begin{equation}
I_1= \int _0 ^1 dz z (1-z) \ln \Big [
\frac {{m_N^*}^2-k^2 z (1-z)}{{m_N}^2 -k^2 z (1-z)} \Big ],
\;\;
I_2= \int _0 ^1 dz \ln \Big [
\frac {{m_N^*}^2-k^2 z (1-z)}{{m_N}^2 -k^2 z (1-z)} \Big ].
\end{equation}
The density dependent part for the self energy
is given as
\begin{equation}
\Pi^D (k_0,{\bfm k} \rightarrow 0)
=-\frac {4 g_{NV}^2}{\pi^2}\int p^2 dp\, F(|{\bfm p}|,m_N^*)
\left [  f_{FD}(\mu ^* ,T)+{\bar f}_{FD}(\mu ^* ,T)\right]
\end{equation}
with
\begin{eqnarray}
F(|{\bfm p}|,m_N^*) &=& \frac {1}{\epsilon^*(p)(4 \epsilon^*(p)^2-k_0^2)}
\Bigg [ \frac {2}{3} (2 |{\bfm p}|^2+3 {m_N^*}^2)+k_0^2 \Big \{ 2 m_N^*
\Big (\frac {\kappa_V}{2 m_N} \Big) \nonumber \\
&& + \frac {2}{3} \Big ( \frac {\kappa_V}{2 m_N}\Big )^2
(|{\bfm p}|^2+3 {m_N^*}^2)\Big \} \Bigg ]
\end{eqnarray}
where $\epsilon ^* (p)=({\bfm p}^2+{m_N^*}^2)^{1/2}$ is the effective
energy for the nucleon.
The effective mass of the vector meson is then obtained by solving
the equation, with $\Pi =\Pi_F +\Pi_D$,
\begin{equation}
k_0^2-m_V^2 + {\rm Re} \Pi (k_0,{\bfm k}=0) =0.
\label {omgrho}
\end{equation}

\subsection {Meson decay properties}

We next proceed to study the vector meson decay widths as modified
due to the effect of vacuum polarisation effects through RHA.
The decay width for the process $\rho \rightarrow \pi \pi$ is
calculated from the imaginary part of the self energy 
and in the rest frame of the $\rho$-meson, it becomes
\begin{equation}
\Gamma _\rho (k_0)=\frac {g_{\rho \pi \pi}^2}{48\pi}
\frac {(k_0^2-4 m_\pi^2)^{3/2}}{k_0^2} \Bigg [
\Big (1+f(\frac {k_0}{2}) \Big )
\Big (1+f(\frac {k_0}{2}) \Big )
-f(\frac {k_0}{2}) f(\frac {k_0}{2}) \Bigg ]
\label {gmrho}
\end{equation}
where, $f(x)=[e^{\beta x}-1]^{-1}$ is the Bose-Einstein distribution
function. The first and the second terms in the above equation 
represent the decay and the formation of the resonance, $\rho$.
The medium effects have been shown to play a very important role
for the $\rho$-meson decay width. In the calculation
for the $\rho$ decay width, the pion has been treated as free,  
i.e. any modification of the pion propagator due to
effects like delta-nucleon hole excitation \cite {asakawa}
have been neglected.
The coupling $g_{\rho \pi \pi}$ is fixed from the decay width
of $\rho$ meson  in vacuum ($\Gamma_\rho$=151 MeV) decaying
into two pions.

For the nucleon-rho couplings, the vector and tensor couplings
as obtained from the N-N forward dispersion relation
\cite{hatsuda1,sourav,grein} are used. 
With the couplings as described above, we
consider the modification of $\omega$ and $\rho$ meson properties in nuclear
matter due to quantum correction effects.

To calculate the decay width for the $\omega$-meson, we
consider the following interaction Lagrangian for the $\omega$ meson
\cite {sakurai,gellmann,bali}
\begin{equation}
{\cal L}_\omega=
\frac {g_{\omega \pi \rho}}{m_\pi}\epsilon _{\mu \nu \alpha  \beta}
\partial ^\mu \omega^\nu \partial ^\alpha {\rho}^{\beta}_i
{\pi_i}+
\frac {g_{\omega 3 \pi}}{m_\pi ^3}\epsilon _{\mu \nu \alpha  \beta}
\epsilon _{ijk}\omega ^\mu \partial ^\nu \pi ^i  \partial ^\alpha \pi ^j
\partial ^\beta \pi ^k.
\label{lomg}
\end {equation}
The decay width of the $\omega$-meson in vacuum is dominated
by the channel $\omega \rightarrow 3 \pi$.
In the medium, the decay width for $\omega \rightarrow 3\pi$
is given as
\begin{eqnarray}
\Gamma _{\omega \rightarrow 3 \pi} & = & \frac {(2\pi)^4}{2 k_0}
\int d^3 {\tilde p}_1   d^3 {\tilde p}_2   d^3 {\tilde p}_3\,
\delta ^{(4)} (P-p_1-p_2-p_3) |M_{fi}|^2
\nonumber \\ && \Big [(1+f(E_1))(1+f(E_2))(1+f(E_3))
-f(E_1)f(E_2)f(E_3) \Big ],
\end{eqnarray}
where $d^3 {\tilde p}_i=\frac {d^3 p_i}{(2\pi)^3 2 E_i}$, $p_i$ and
$E_i$'s are 4-momenta and energies for the pions, and $f(E_i)$'s
are their thermal distributions. The matrix element $M_{fi}$ has
contributions from the channels
$\omega \rightarrow \rho \pi \rightarrow 3\pi$ (described by the first
term in (\ref {lomg})) and the direct decay $\omega \rightarrow 3\pi$
resulting from the contact interaction (second term in (\ref {lomg}))
\cite{bali,weisezp,kaymak}.
For the $\omega \rho \pi$ coupling we take the value $g_{\omega \rho \pi}$
=2 which is compatible to the vacuum decay width 
$\omega \rightarrow \pi \gamma$ \cite{sourav}.
We fix the point interaction coupling $g_ {\omega 3 \pi}$
by fitting the partial decay width $\omega \rightarrow 3 \pi$ in vacuum
(7.49 MeV) to be 0.24. The contribution arising from the direct decay
turns out to be marginal, being of the order of upto 5\% of the
total decay width for $\omega \rightarrow 3\pi$.

With the modifications of the vector meson masses in the hot and dense
medium, a new channel becomes accessible, the decay mode
$\omega \rightarrow \rho \pi$ for $m^*_\omega > m^*_\rho +m_\pi$.
This has been taken into account in the present investigation.

\section{Results and Discussions}
\label{results}
\begin{figure}
\begin{center}
\centerline{\parbox[b]{8cm}{
\includegraphics[width=9.2cm,height=9cm]{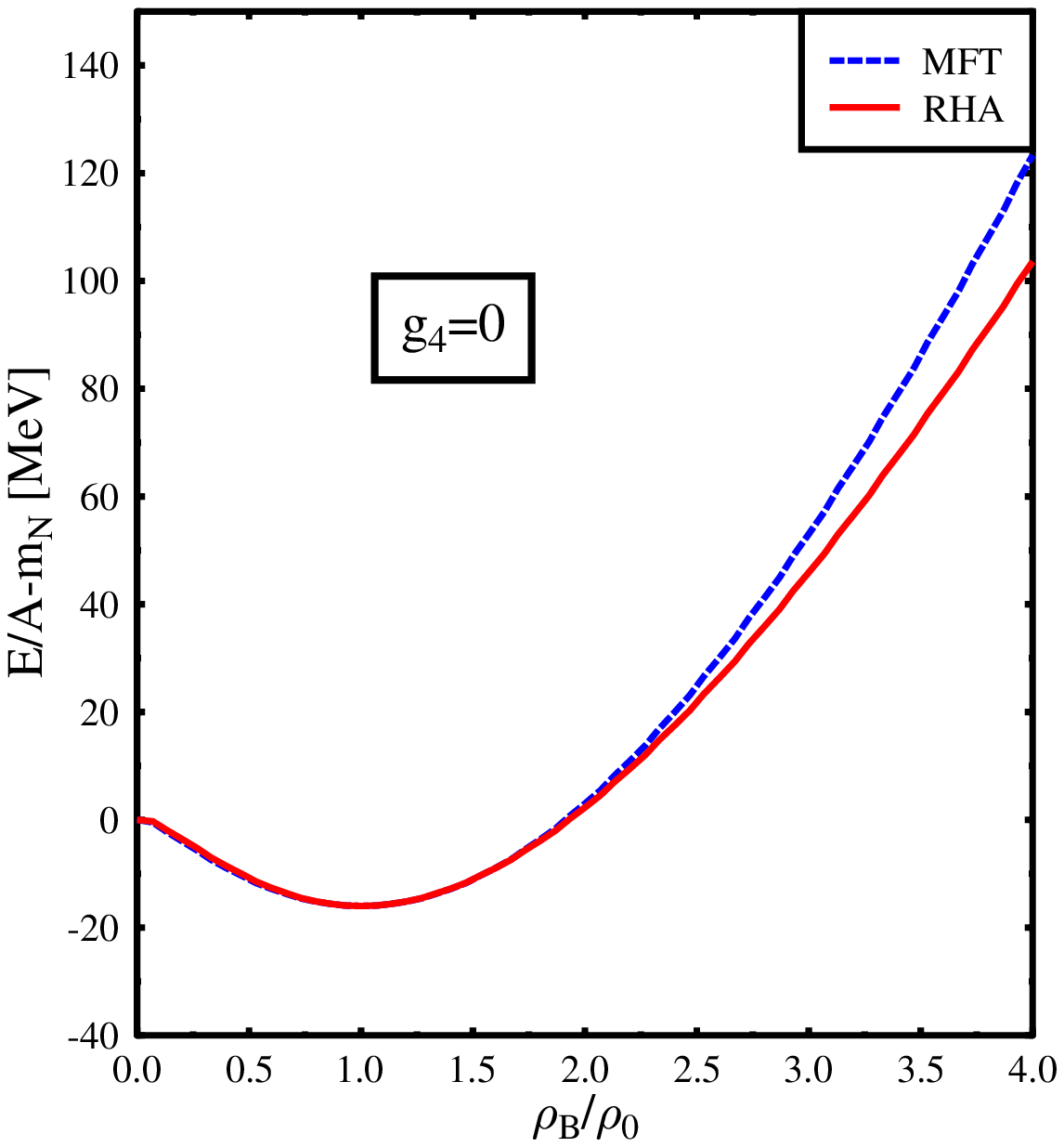}}
\parbox[b]{8cm}{
\includegraphics[width=9.2cm,height=9cm]{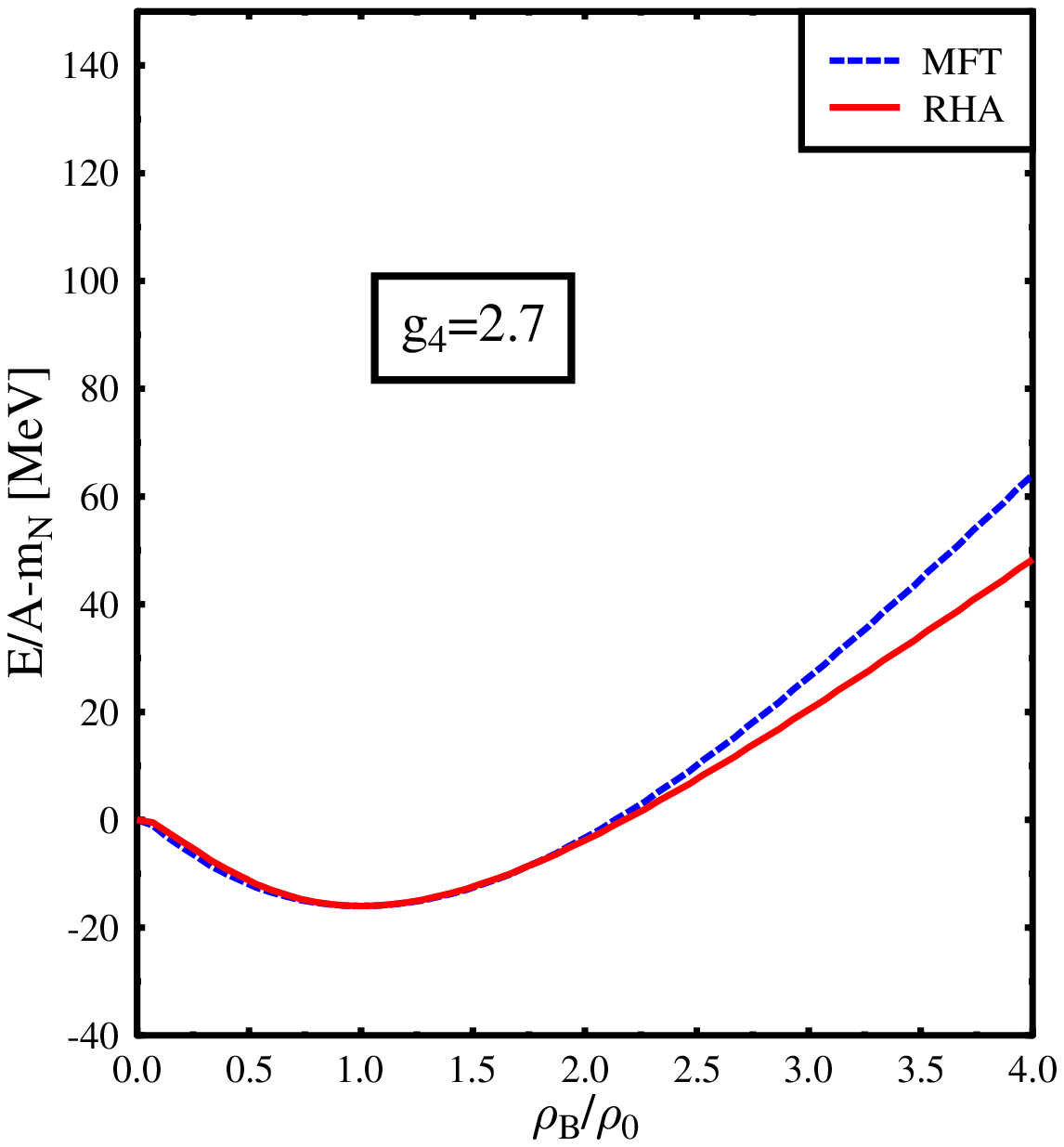}}}
\vspace{-1cm}
\caption{
\label{eos}
Binding energy per particle as a function of density in the Mean-Field
and in the Hartree-Approximation.
}
\end{center}
\end{figure}

\begin{figure}
\begin{center}
\centerline{\parbox[b]{8cm}{
\includegraphics[width=9.2cm,height=9cm]{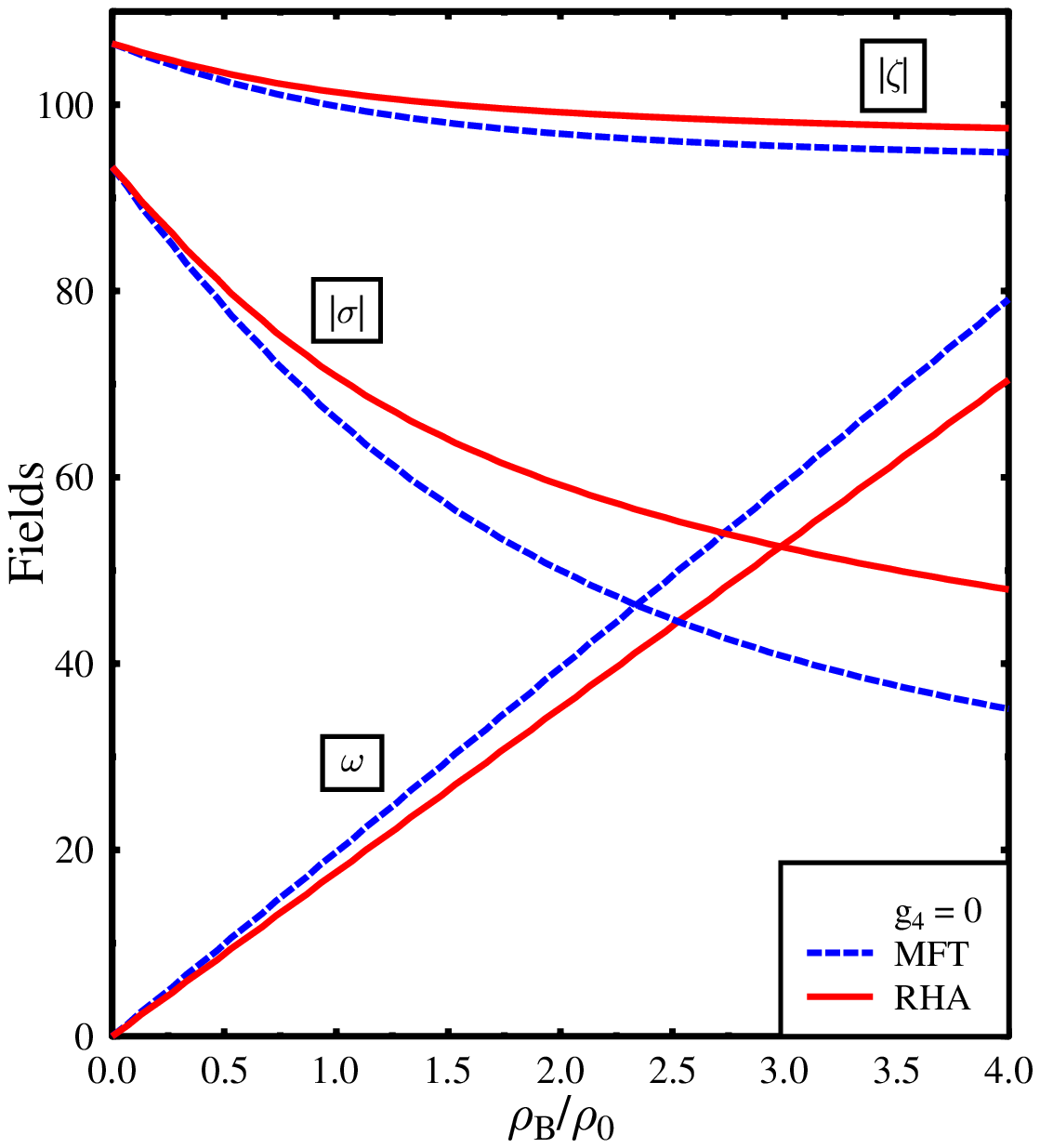}}
\parbox[b]{8cm}{
\includegraphics[width=9.2cm,height=9cm]{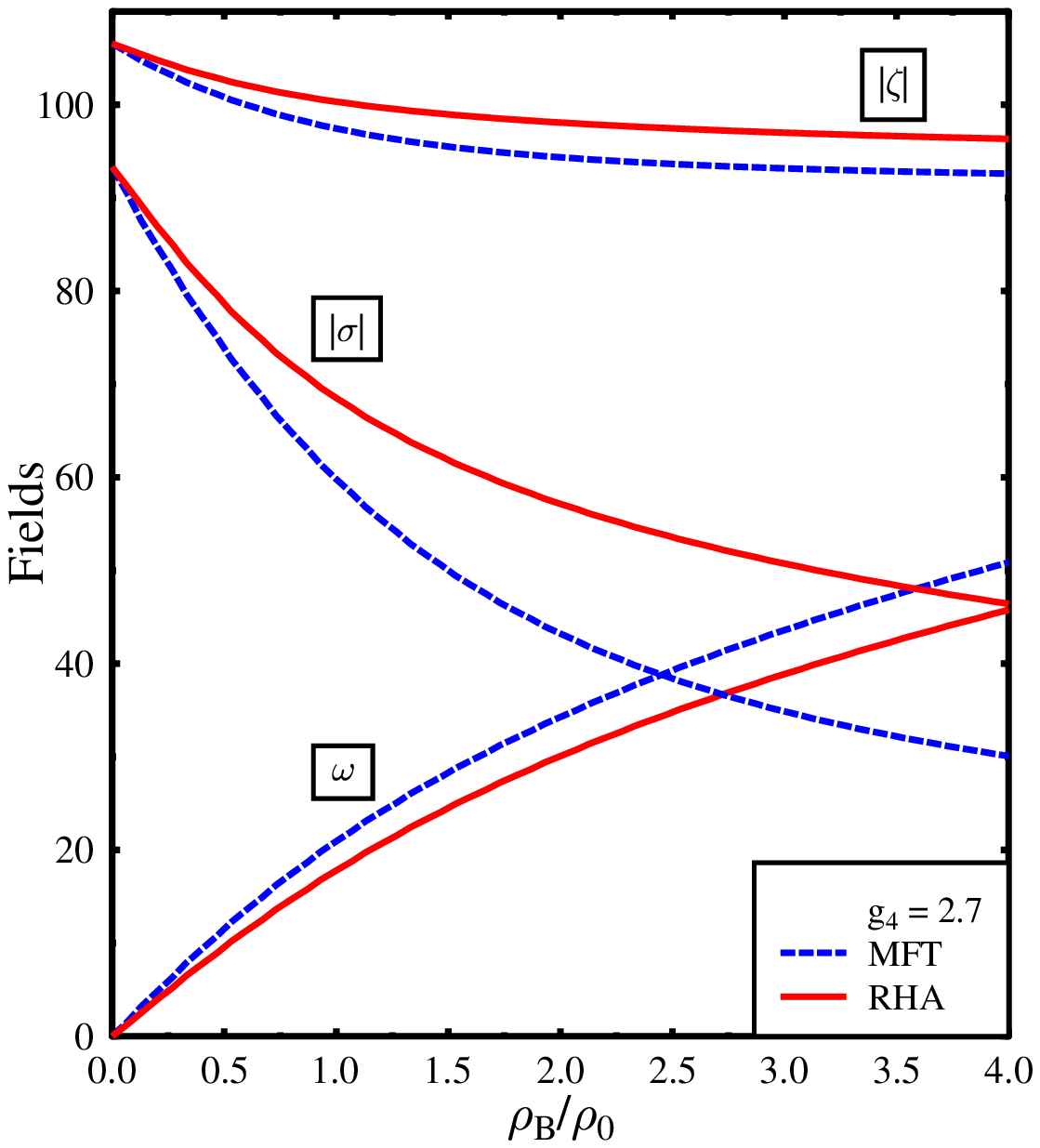}}}
\vspace{-1cm}
\caption{
\label{fields}
Scalar fields $\sigma$, $\zeta$ and vector field $\omega$
as a function of density in the Mean-Field and in the
Hartree-Approximation.
}
\end{center}
\end{figure}

We shall now discuss the results of the present investigation:
the nucleon properties as modified due to the Dirac sea contributions
through the relativistic Hartree approximation and their 
effects on vector meson properties in the dense hadronic matter.
Figure \ref{eos} shows the equation of state in the Mean-Field- and in
the Hartree-approximation with and without quartic self-interaction
for the $\omega$-field. 
In both cases we observe that the additional 
terms resulting from the Hartree approximation lead to a softening of the
equation of state at higher densities. However, the compressibilty in
the relativistic Hartree-approximation is higher than the mean-field
value, as shown in table \ref{parmfhart}.
Furthermore, the influence of the finite value for the quartic
$\omega$ coupling, $g_4$ is clearly visible. In this case the 
compressibility at
nuclear saturation is strongly reduced (table \ref{parmfhart}). Also,
the resulting equation of state is much softer in particular
at higher densities. The reason for this can be seen from figure
\ref{fields}. The vector field $\omega$, which causes the
repulsion in the system, rises much more steeply as a function of density 
for $g_4=0$ than for the case of $g_4=2.7$, because the quartic
self-interaction attenuates the $\omega$-field. 

\begin{figure}
\begin{center}
\centerline{\parbox[b]{8cm}{
\includegraphics[width=9.2cm,height=9cm]{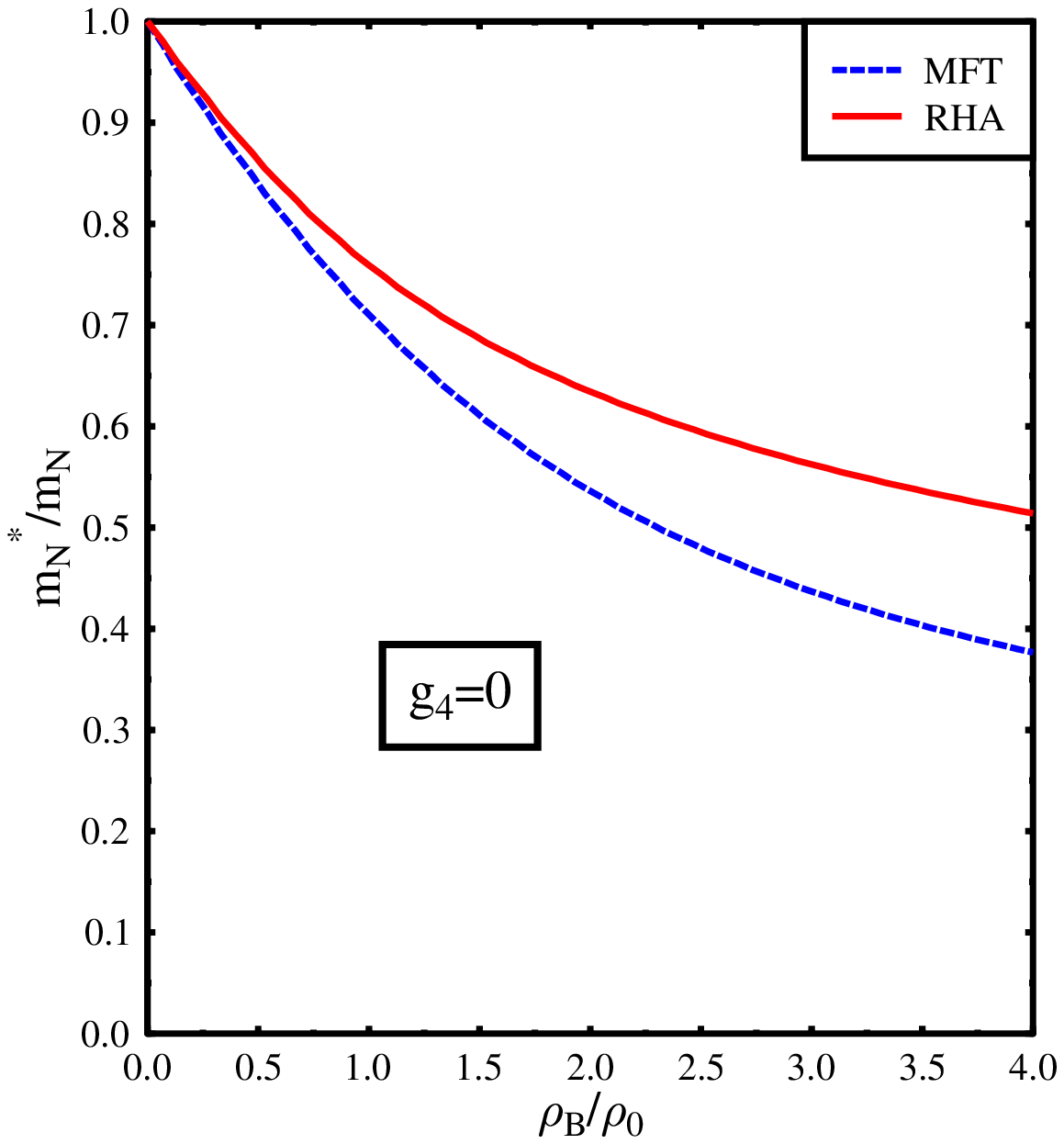}}
\parbox[b]{8cm}{
\includegraphics[width=9.2cm,height=9cm]{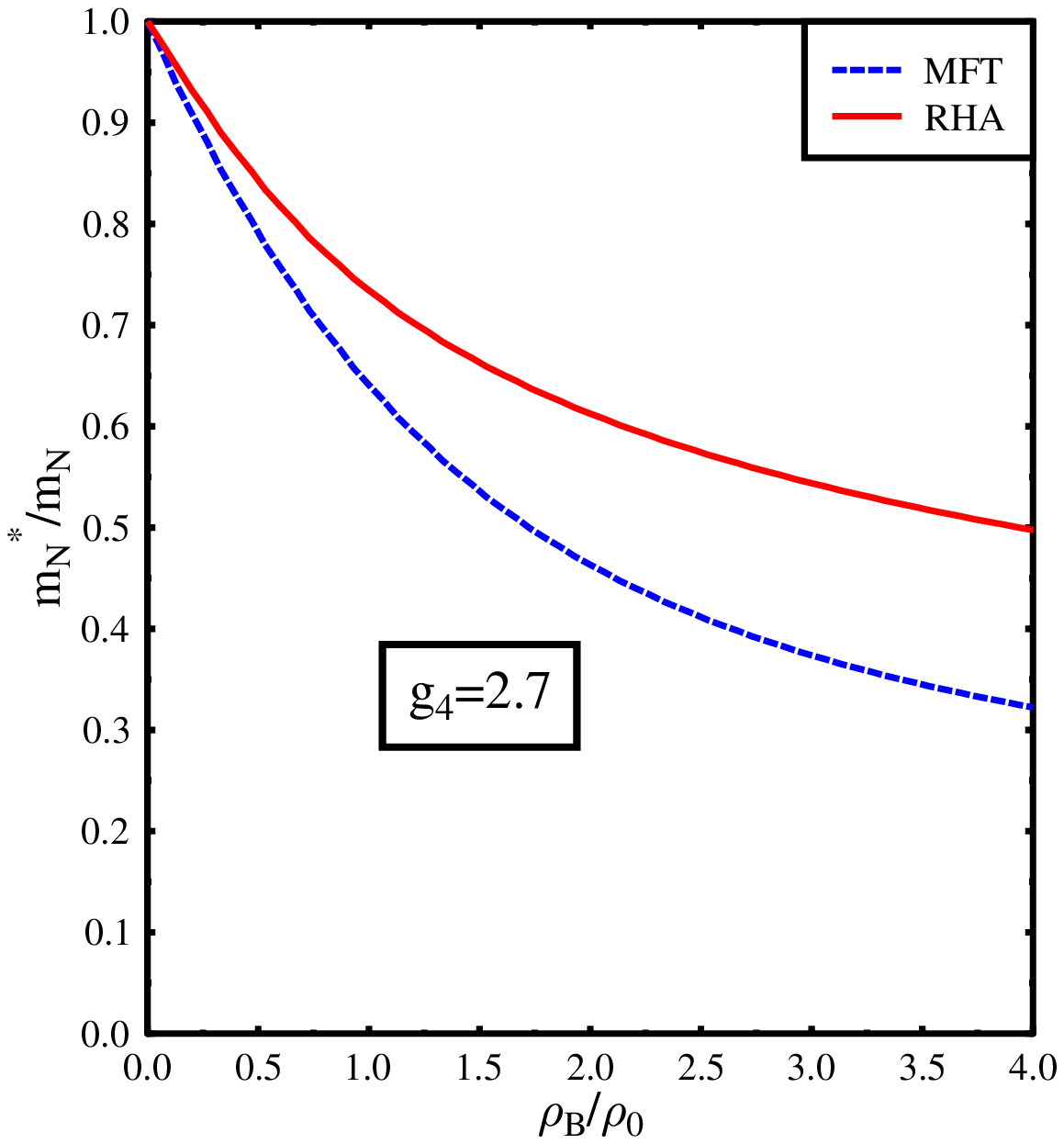}}}
\vspace{-1cm}
\caption{
\label{effmn}
Effective nucleon mass as a function of density in the Mean-Field
and in the Hartree-Approximation.
}
\end{center}
\end{figure}

\begin{figure}
\begin{center}
\centerline{\parbox[b]{8cm}{
\includegraphics[width=9.2cm,height=9cm]{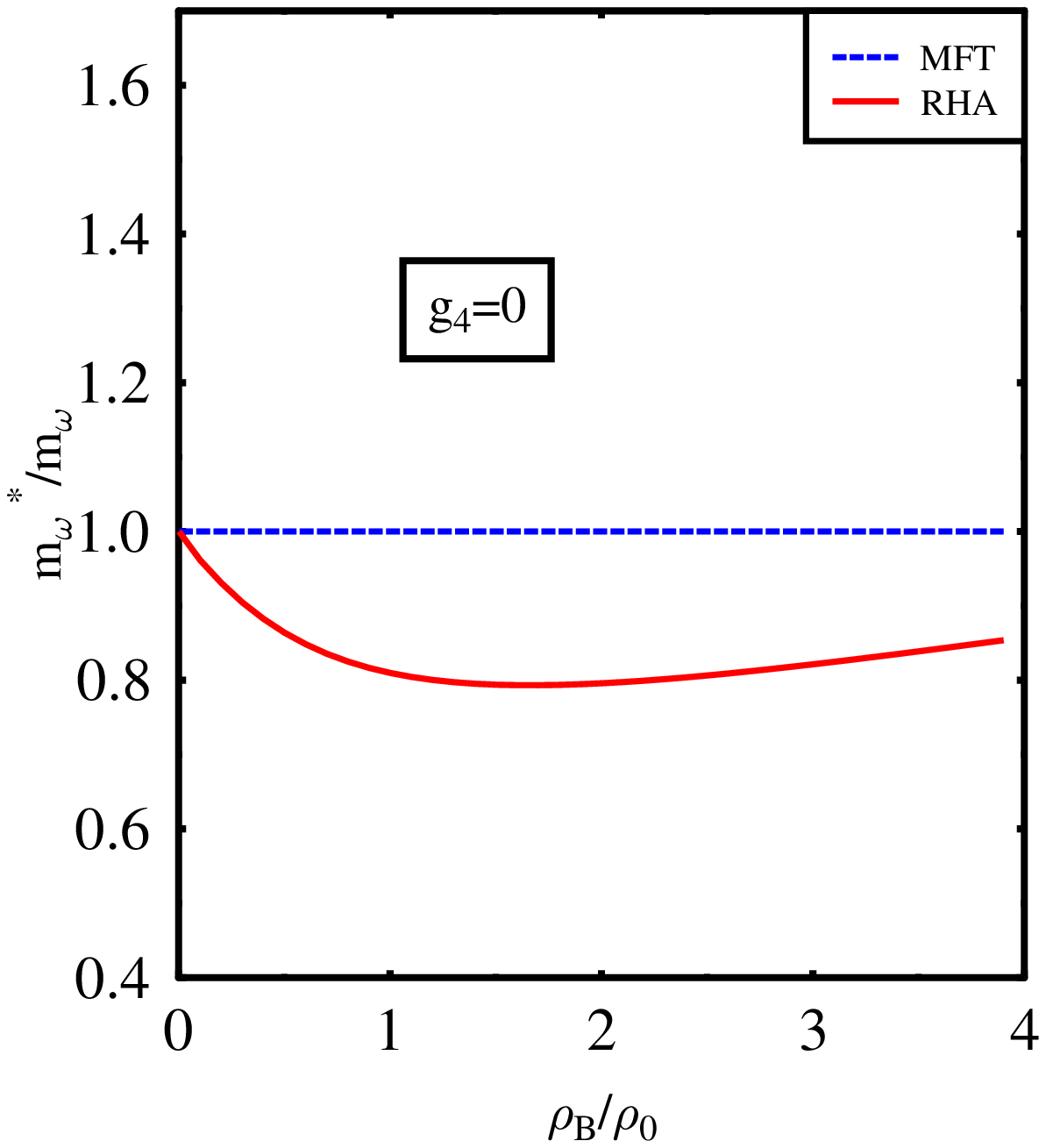}}
\parbox[b]{8cm}{
\includegraphics[width=9.2cm,height=9cm]{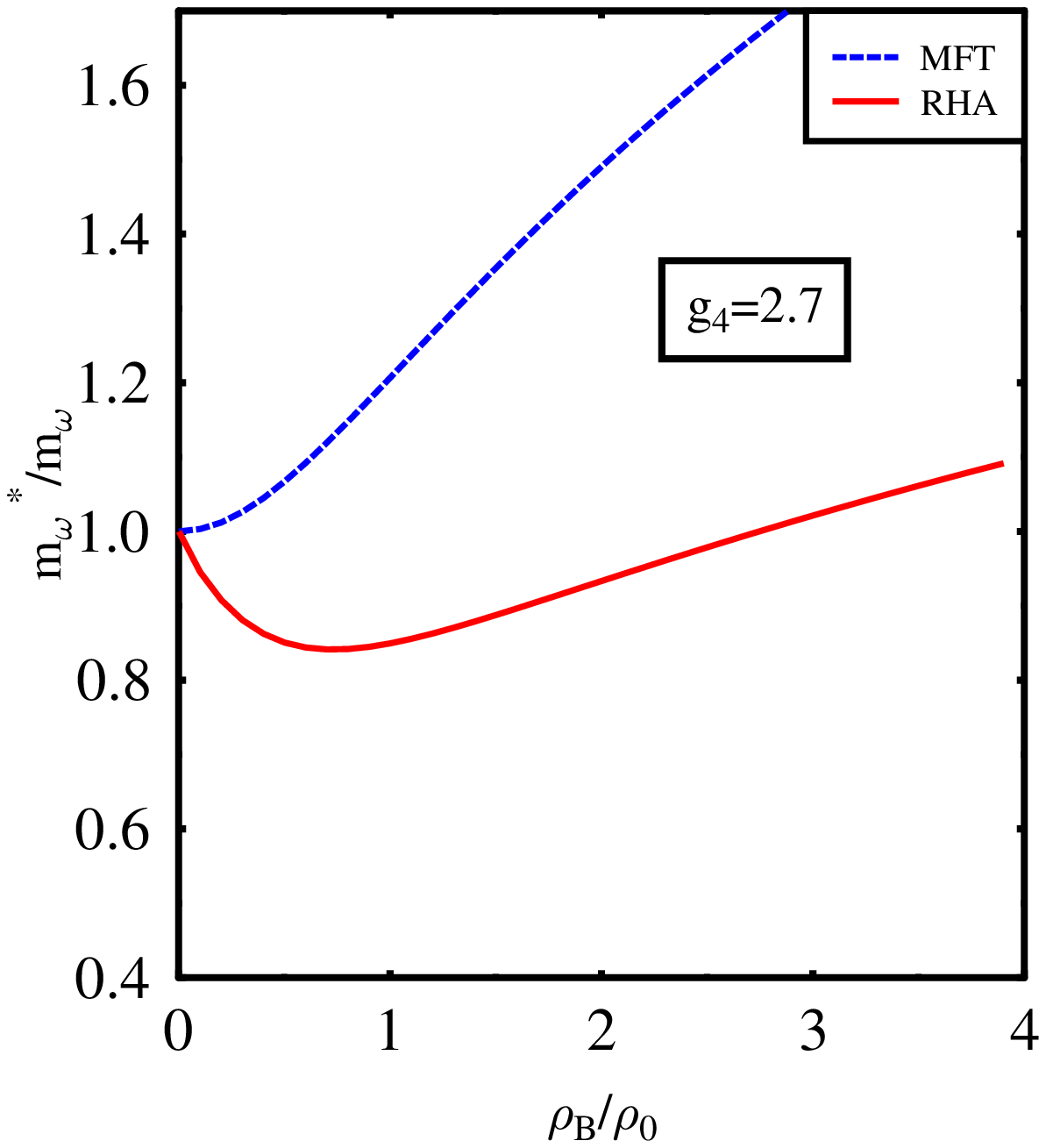}}}
\vspace{-1cm}
\caption{
\label{momgfig}
Effective $\omega$ meson mass in the mean field approximation and
including the Hartree contributions. Left: no quartic vector-meson
interaction. Right: Including $\omega^4$ interaction. 
There is significant drop
of the vector meson mass due to the Dirac sea effect, which is not seen
in the mean field approximation.
}
\end{center}
\end{figure}

The effective nucleon mass for the different cases is depicted in
figure \ref{effmn}. Here the RHA predicts higher nucleon masses than
the Mean-Field case. At higher densities these contributions become
increasingly important.
This is also reflected in the density dependence of the non-strange
$\sigma$ field, showing a considerable increase due to the 
Hartree contributions (figure\ref{fields}).
In contrast, the strange condensate, $\zeta$, which does not couple to the
nucleons, takes only slightly lower values in the MF-case. 
The in-medium properties of the vector mesons
are modified due to the vacuum polarization effects.
The nucleon-$\omega$ vector coupling, $g_{N \omega}$,
is calculated from the nuclear matter saturation properties.
As already stated, the N-$\omega$ tensor coupling is neglected.
Figure \ref{momgfig} shows the resulting 
modification of the $\omega$--meson mass
in the Hartree approximation as compared to the mean field case.
For $g_4=0$, the $\omega$ mass has no density dependence, because of  
the frozen glueball approximation.
In contrast, a strong
reduction due to the Dirac sea polarization is found for densities up
to around normal nuclear matter density. At higher  densities,
the Fermi polarization part of the $\omega$-self-energy starts to become 
important, leading to an increase in the mass.
A similar behaviour has been observed in the Walecka model 
\cite{hatsuda,hatsuda1,jeans}.
The quartic term in the $\omega$-field considerably enhances the
$\omega$ mass with increasing density. Thus in the mean field case, 
the mass rises monotonically. For the Hartree
approximation a decrease of the $\omega$ mass 
for small densities can still be found. But at higher densities the 
contribution from the quartic term
becomes more important and leads to an increase of the in-medium mass.

\begin{figure}
\begin{center}
\centerline{\parbox[b]{8cm}{
\includegraphics[width=9.2cm,height=9cm]{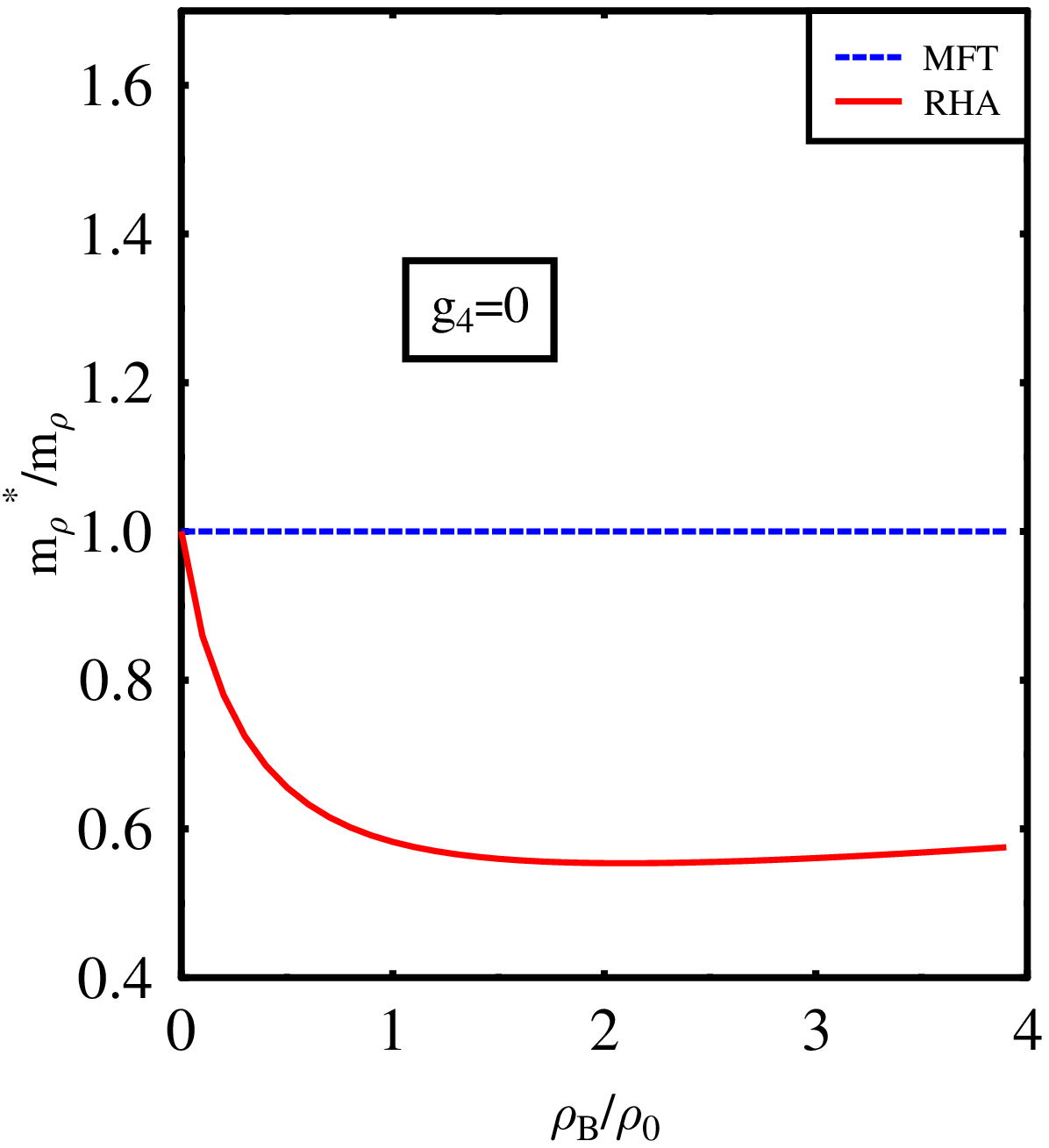}}
\parbox[b]{8cm}{
\includegraphics[width=9.2cm,height=9cm]{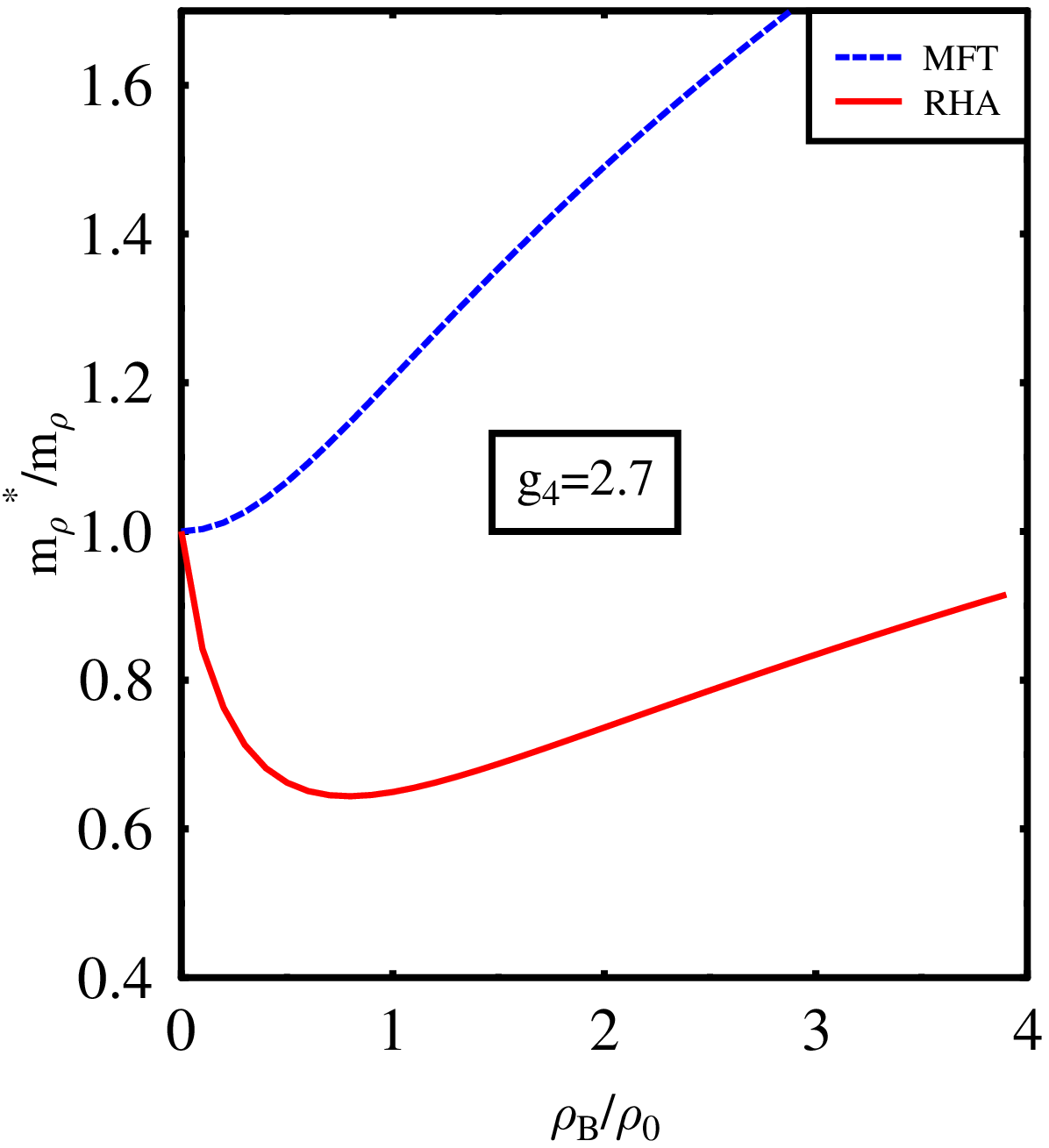}}}
\vspace{-1cm}
\caption{
\label{mrhofig}
Effective $\rho$ meson mass without and with the
Hartree contributions, with the nucleon-rho vector and tensor couplings,
as fitted from the NN scattering data ($g_{N\rho}$=2.63, $\kappa_\rho=6.1$).
The Hartree approximation gives rise to the decrease of the
$\rho$ mass in the medium.
}
\end{center}
\end{figure}

In figure \ref{mrhofig}, we illustrate the medium modification for
the $\rho$ meson mass with the vector and tensor couplings
to the nucleons being fixed from the NN forward dispersion relation
\cite{hatsuda1,sourav,grein}. The values
for these couplings are given as $g_{N \rho}^2/4\pi$=0.55
and $\kappa_\rho$=6.1. We notice that the decrease in the
$\rho$ meson with increasing density is much sharper than
that of the $\omega$ meson. Such a behaviour of the $\rho$ meson
undergoing a much larger medium modification was also observed
earlier \cite {sourav} within the relativistic Hartree approximation
in the Walecka model. This indicates that the tensor
coupling, which is negligible for
the $\omega$ meson, plays a significant role for the $\rho$ meson.

\begin{figure}
\begin{center}
\centerline{\parbox[b]{8cm}{
\includegraphics[width=9.2cm,height=9cm]{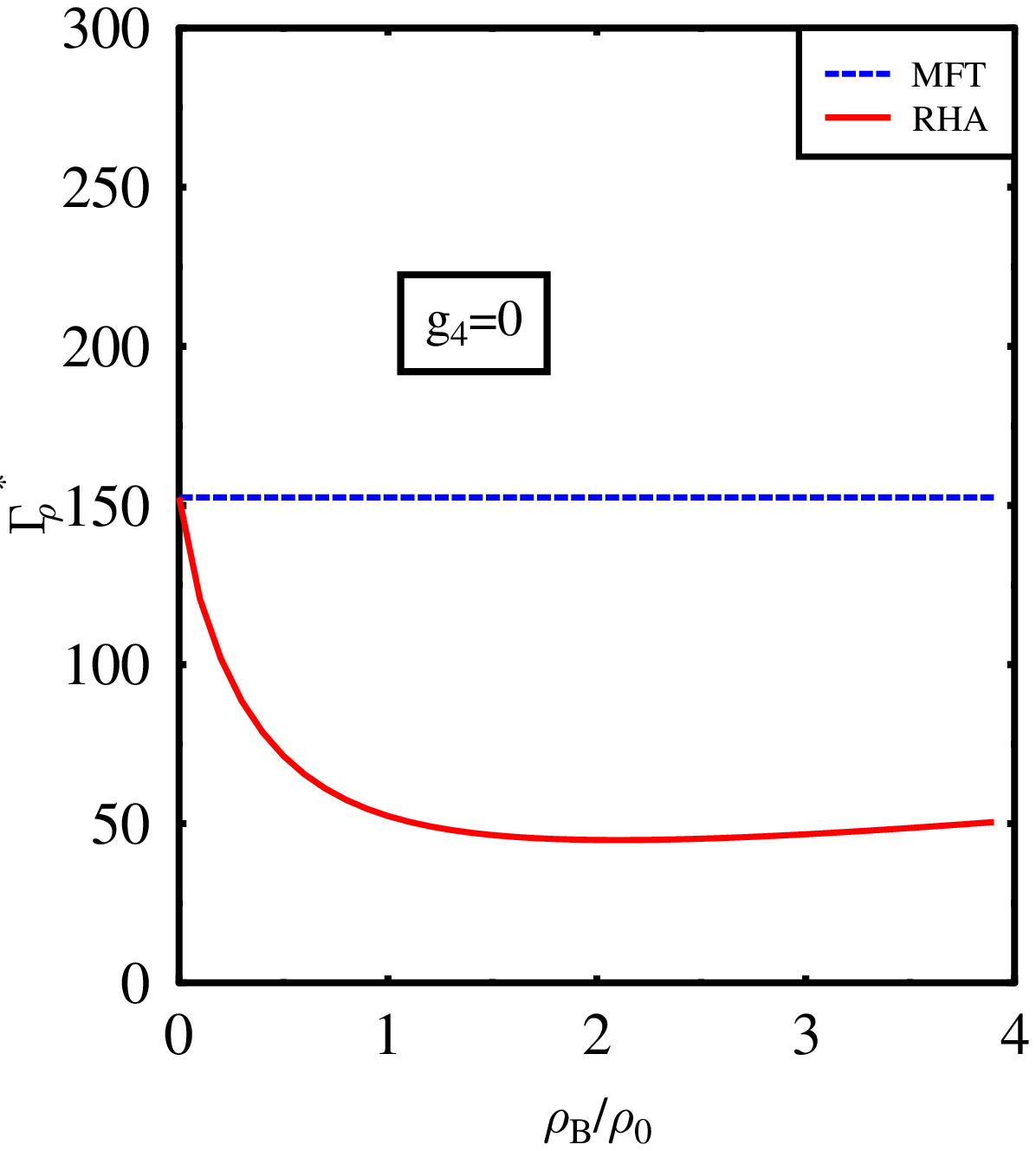}}
\parbox[b]{8cm}{
\includegraphics[width=9.2cm,height=9cm]{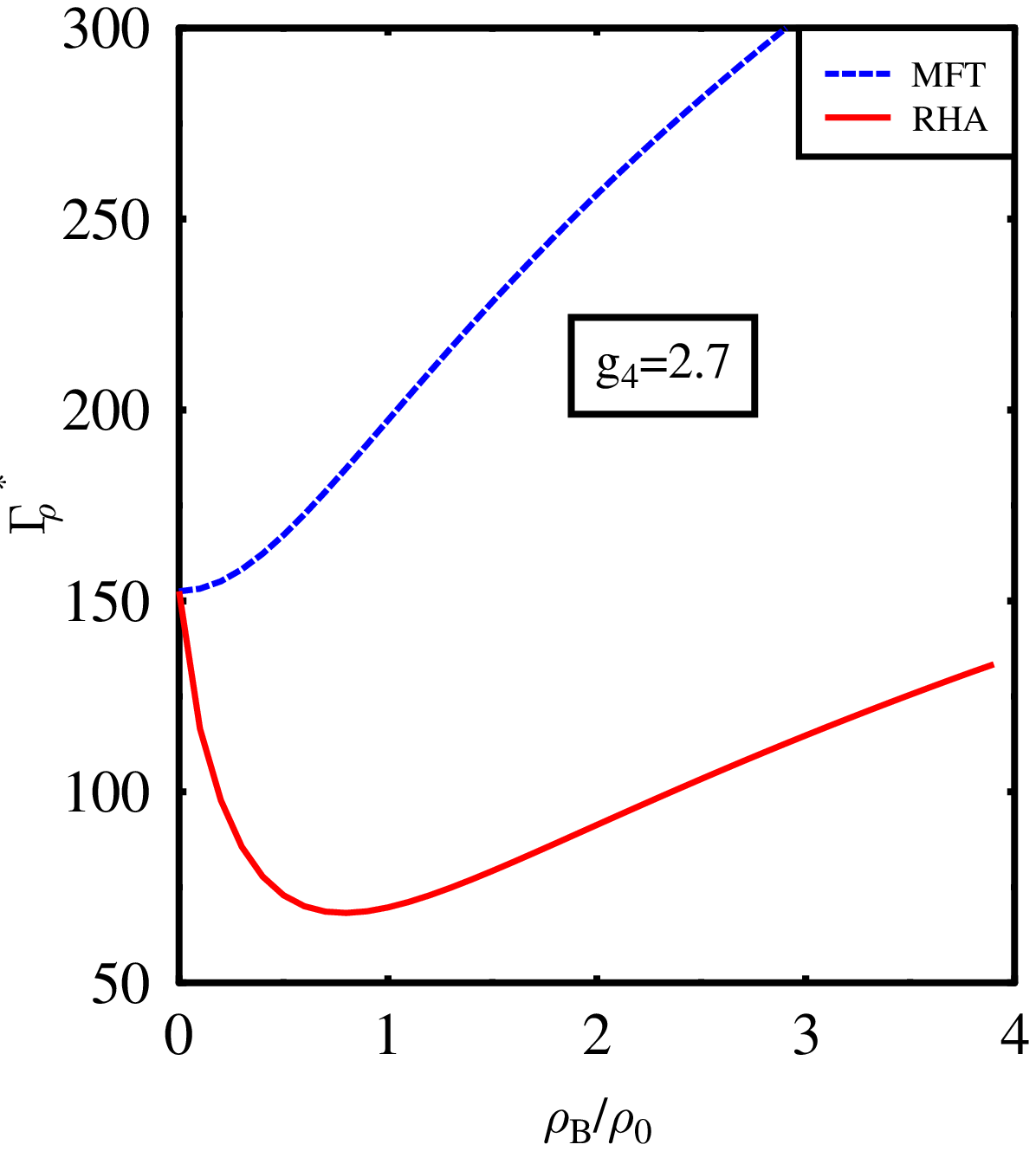}}}
\vspace{-1cm}
\caption{
\label{gmrhofig}
Decay width of $\rho$ meson in the absence and presence of the
Dirac sea effect with the couplings fitted from the NN scattering data.
}
\end{center}
\end{figure}

The in-medium decay width for $\rho \rightarrow \pi\pi$, 
$\Gamma_\rho^\ast$ reflects the
behaviour of the in-medium $\rho$ mass. 
This is because in the present work
only the case $T=0$ is considered, and so there is no Bose-enhancement
effect. 
Therefore in the absence of the quartic vector-meson interaction,
the significant drop of the mass of the $\rho$ meson in the medium
leads to a decrease of 
the $\rho$ decay width. This is shown in figure \ref{gmrhofig}.
Since the quartic self-interaction yields an increase in the
mass at higher densities, 
it leads to an increase of the $\rho$ decay width.

\begin{figure}
\begin{center}
\centerline{\parbox[b]{8cm}{
\includegraphics[width=9.2cm,height=9cm]{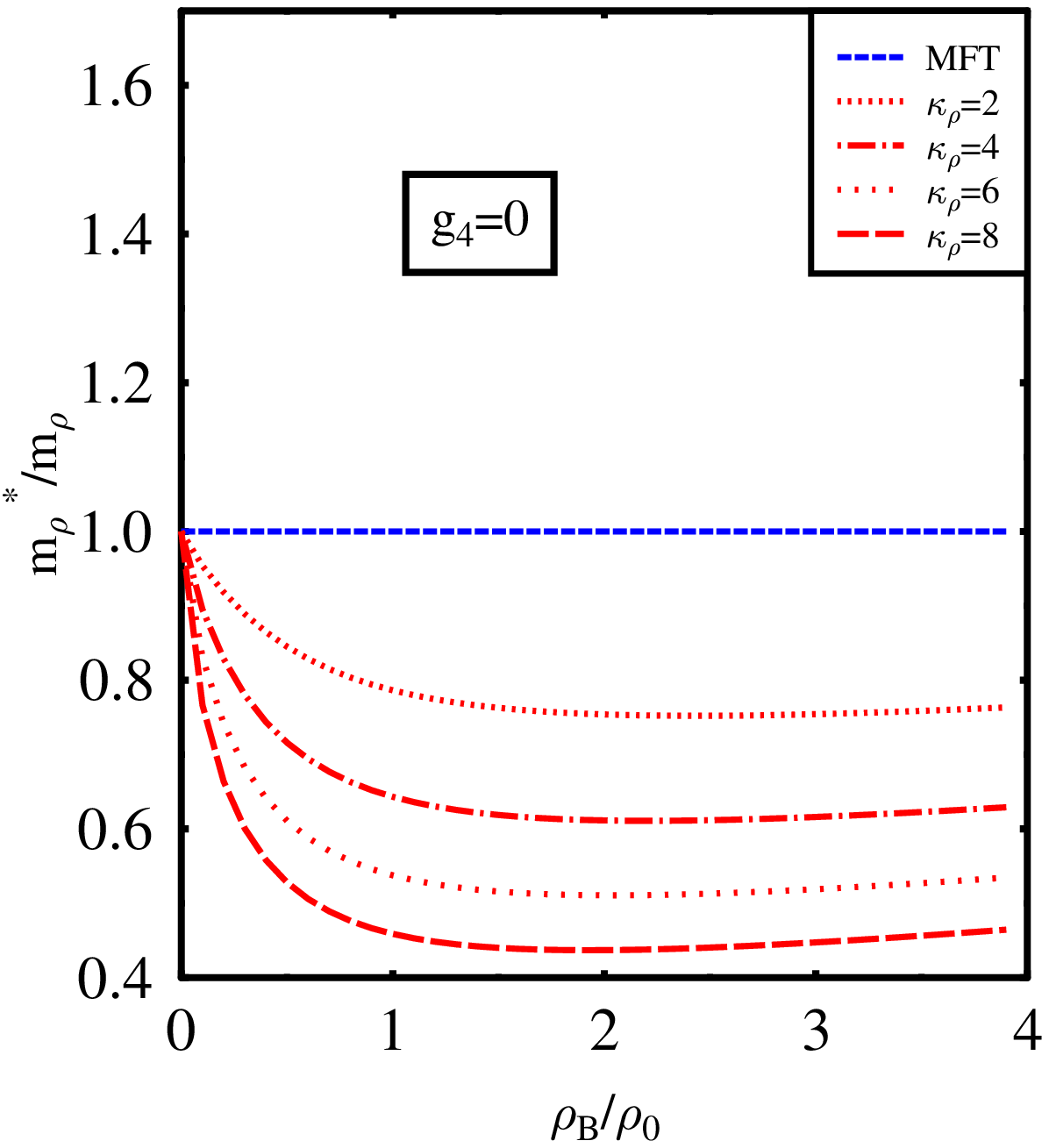}}
\parbox[b]{8cm}{
\includegraphics[width=9.2cm,height=9cm]{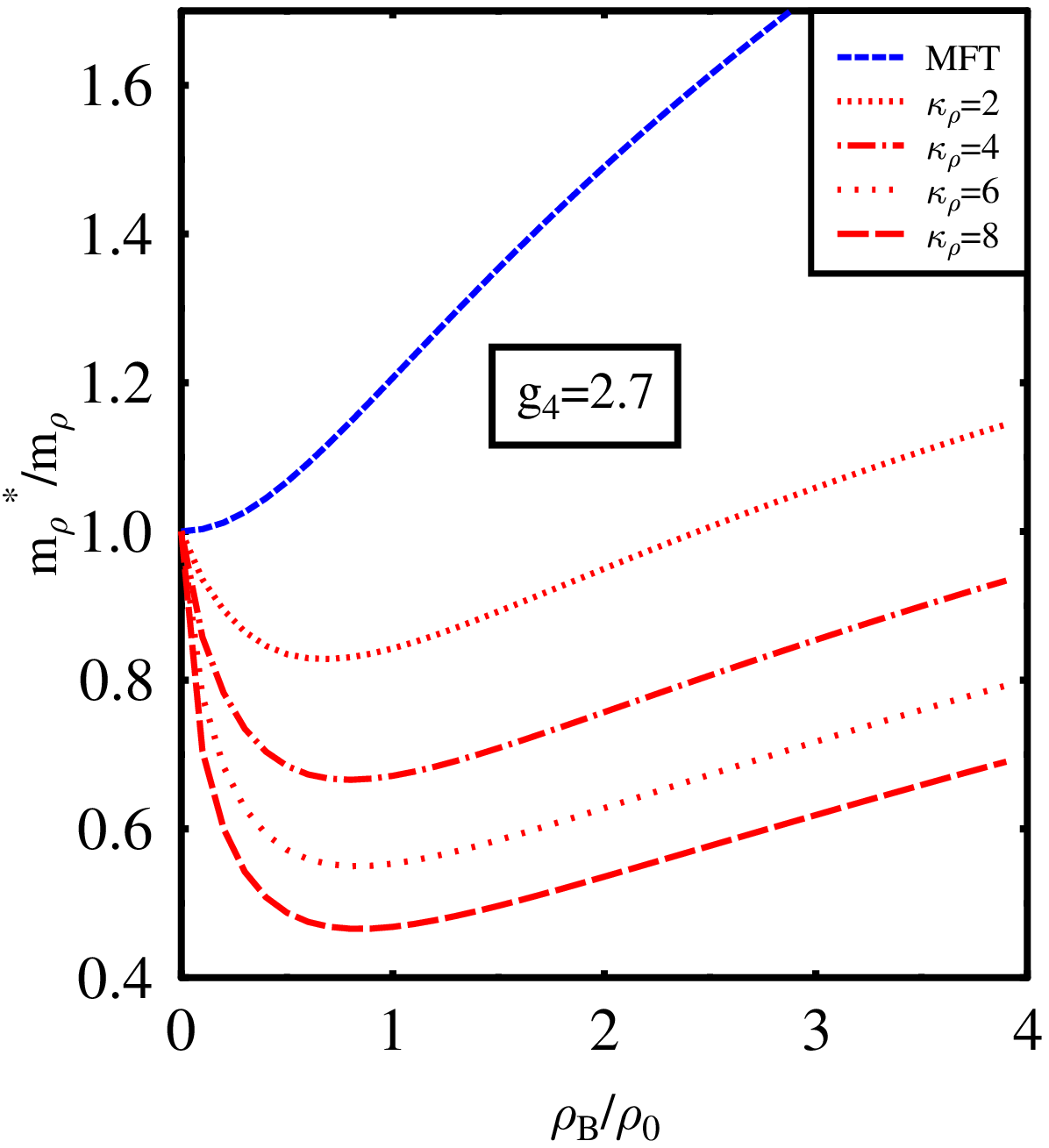}}}
\vspace{-1cm}
\caption{
\label{mrhokfig}
Effective $\rho$ meson mass without and with the
Hartree contributions, with the nucleon-rho vector coupling,
$g_{N\rho}$, as from the chiral model, which is compatible with
the symmetry energy. Since we do not know the medium dependent
tensor coupling, $\kappa_\rho$, it is taken
as a parameter. The Hartree approximation gives rise to the
decrease of the $\rho$ mass in the medium, which is seen to be
quite sensitive to the nucloen- rho tensor coupling.
}
\end{center}
\end{figure}

In the previous calculations, the $\rho$-N coupling strengths were 
used as determined from 
the NN forward scattering data \cite {grein}.
Now we consider the mass modification for the $\rho$ meson,
with the nucleon $\rho$ coupling,
$g_{N\rho}$ as determined from the symmetry relations 
(table \ref{parmfhart}).
The symmetry energy coefficient $a_{sym}$ is
given as \cite {sym}
\begin{equation}
 a_{sym}= \frac {1}{2} \,
\left[ \frac {\partial^{2}}{\partial t^{2}}\,\left(\frac{\epsilon}{\rho}\right)\right]_{t=0} 
\, ,
\end{equation}
where $t=\frac{\rho_n - \rho_p}{\rho_0}$.
The resulting values for the symmetry energy for the different cases are
shown in table \ref{parmfhart}.
They are compatible with the experiment. 
We take the tensor coupling as a parameter in our calculations
since this coupling cannot be fixed from infinite nuclear matter properties.
However, it influences the properties of finite nuclei.
The resulting in-medium mass of the $\rho$ meson 
is plotted in figure \ref{mrhokfig}
as a function of baryon density $\rho_B/\rho_0$.
It is observed that the $\rho$ meson mass has a strong dependence
on the tensor coupling. In the Hartree approximation 
the $\rho$-nucleon vector coupling does
not differ too much in the two cases, i.e. depending on whether 
 it is obtained fromm NN
scattering data or from the symmetry relations. Thus 
we find a similar 
behaviour for the $\rho$ mass, if in the latter case we choose 
$\kappa_\rho = 6$, i.e. close to the value from scattering data.

\begin{figure}
\begin{center}
\centerline{\parbox[b]{8cm}{
\includegraphics[width=9.2cm,height=9cm]{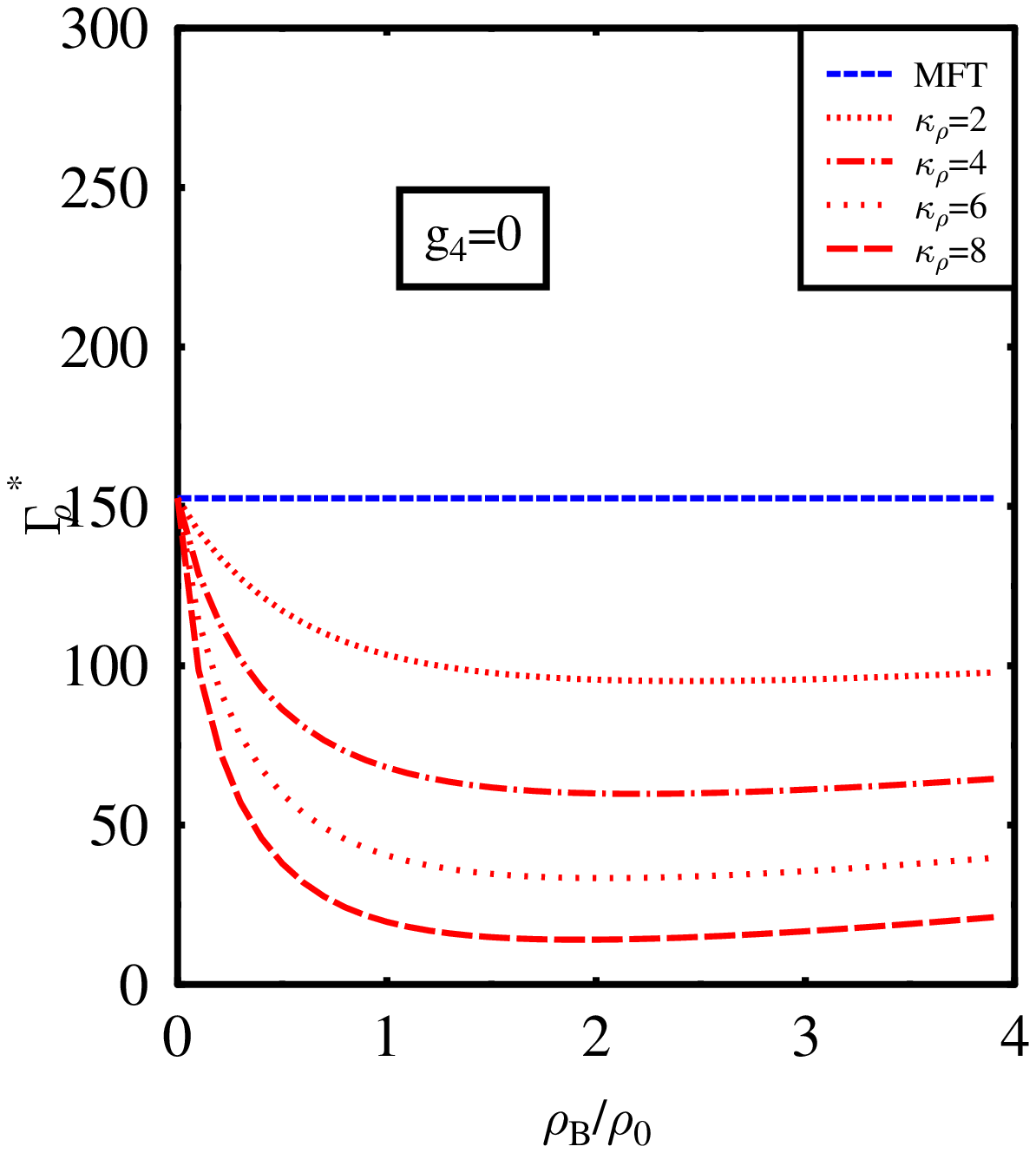}}
\parbox[b]{8cm}{
\includegraphics[width=9.2cm,height=9cm]{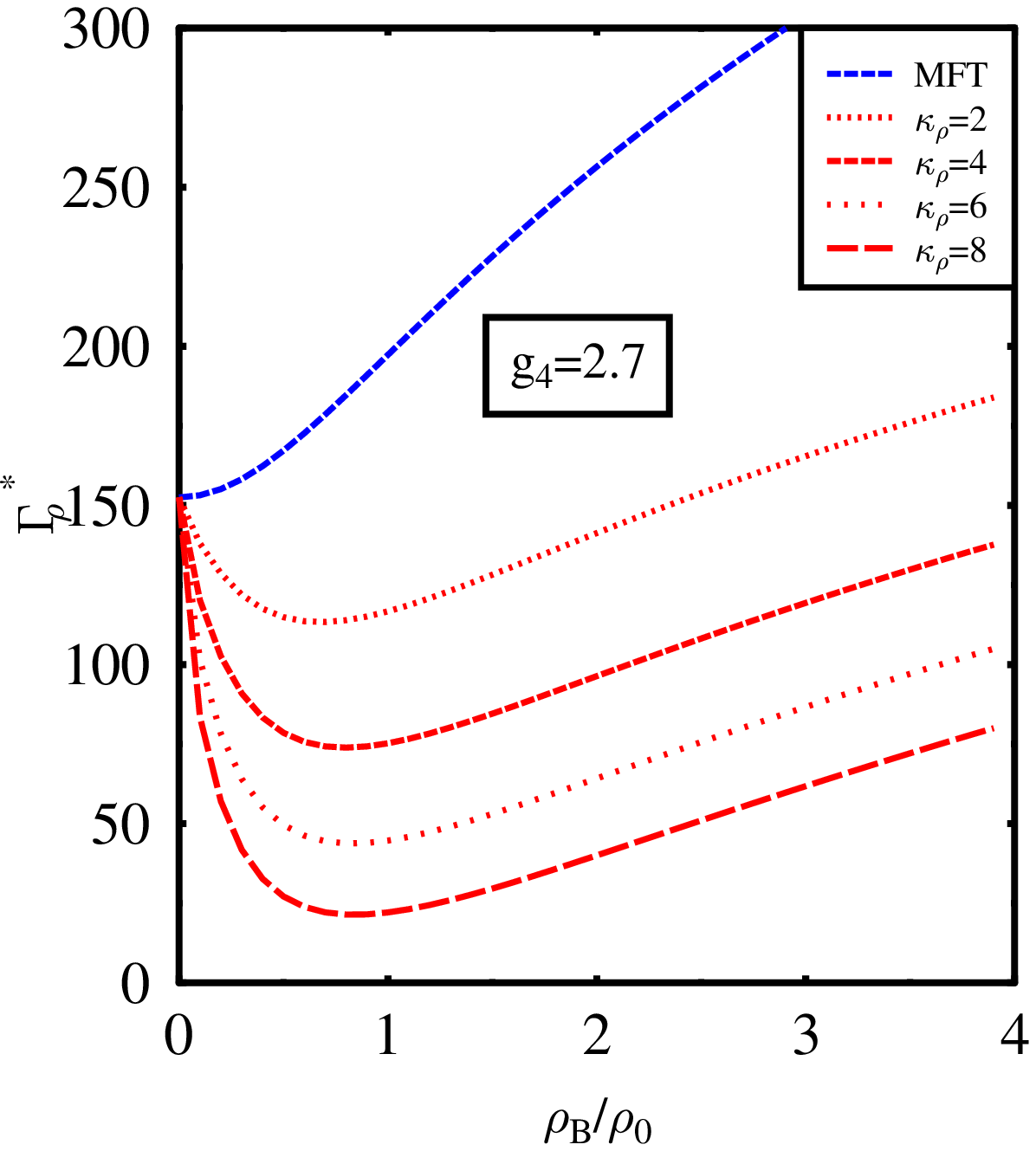}}}
\vspace{-1cm}
\caption{
\label{gmrhokfig}
Decay width of $\rho$ meson in the absence and presence of the
Dirac sea effect with the nucleon-rho vector coupling,
$g_{\rho N}$, as from the chiral model, which is compatible with
the symmetry energy. The tensor coupling, $\kappa_\rho$ is taken
as a parameter.
}
\end{center}
\end{figure}

Figure \ref{gmrhokfig} shows the decay width for the $\rho$ meson
when we take the $N\rho$ vector coupling as determined from symmetry 
relations and the tensor coupling taken as a parameter.

\begin{figure}
\begin{center}
\centerline{\parbox[b]{8cm}{
\includegraphics[width=9.2cm,height=9cm]{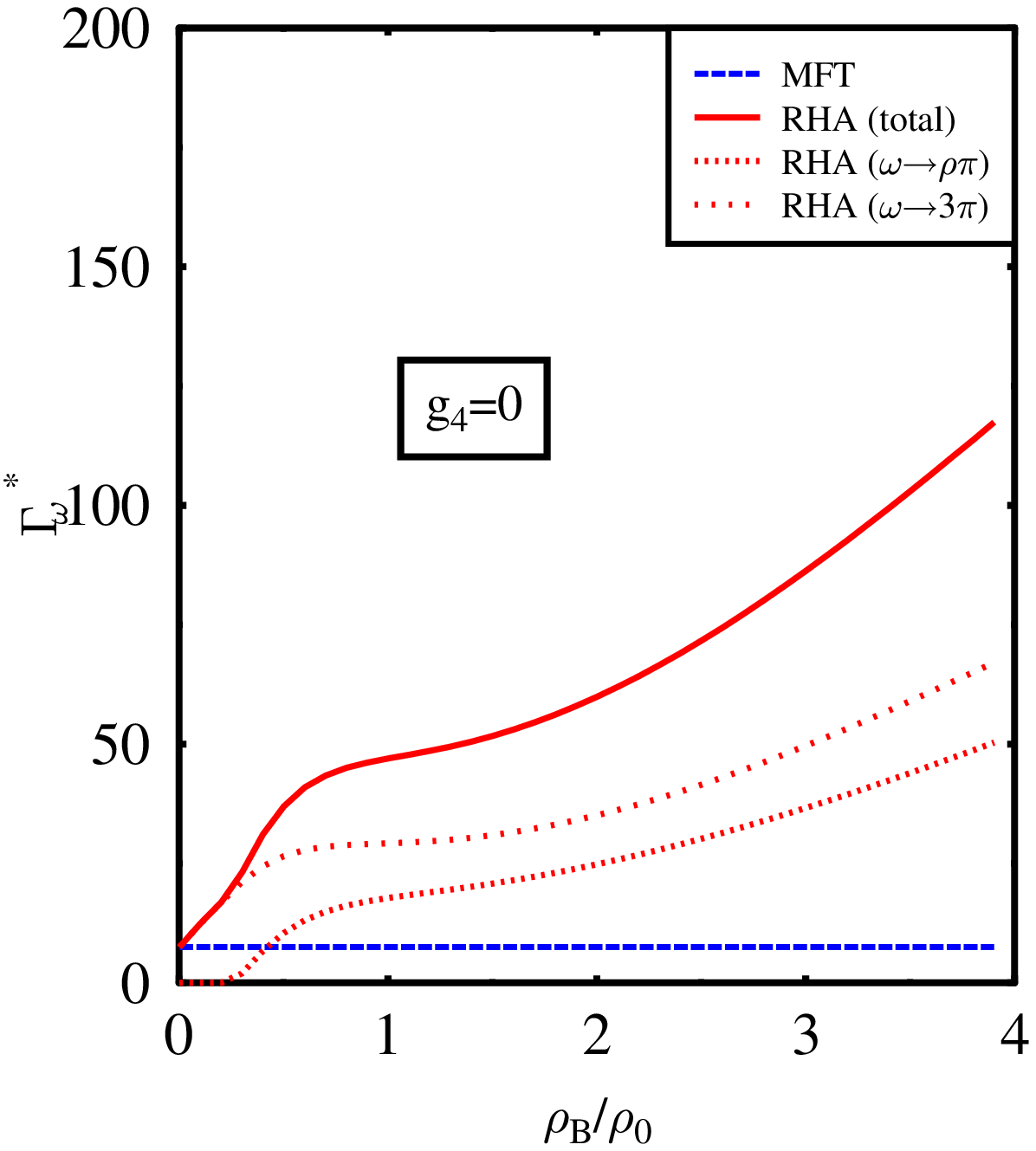}}
\parbox[b]{8cm}{
\includegraphics[width=9.2cm,height=9cm]{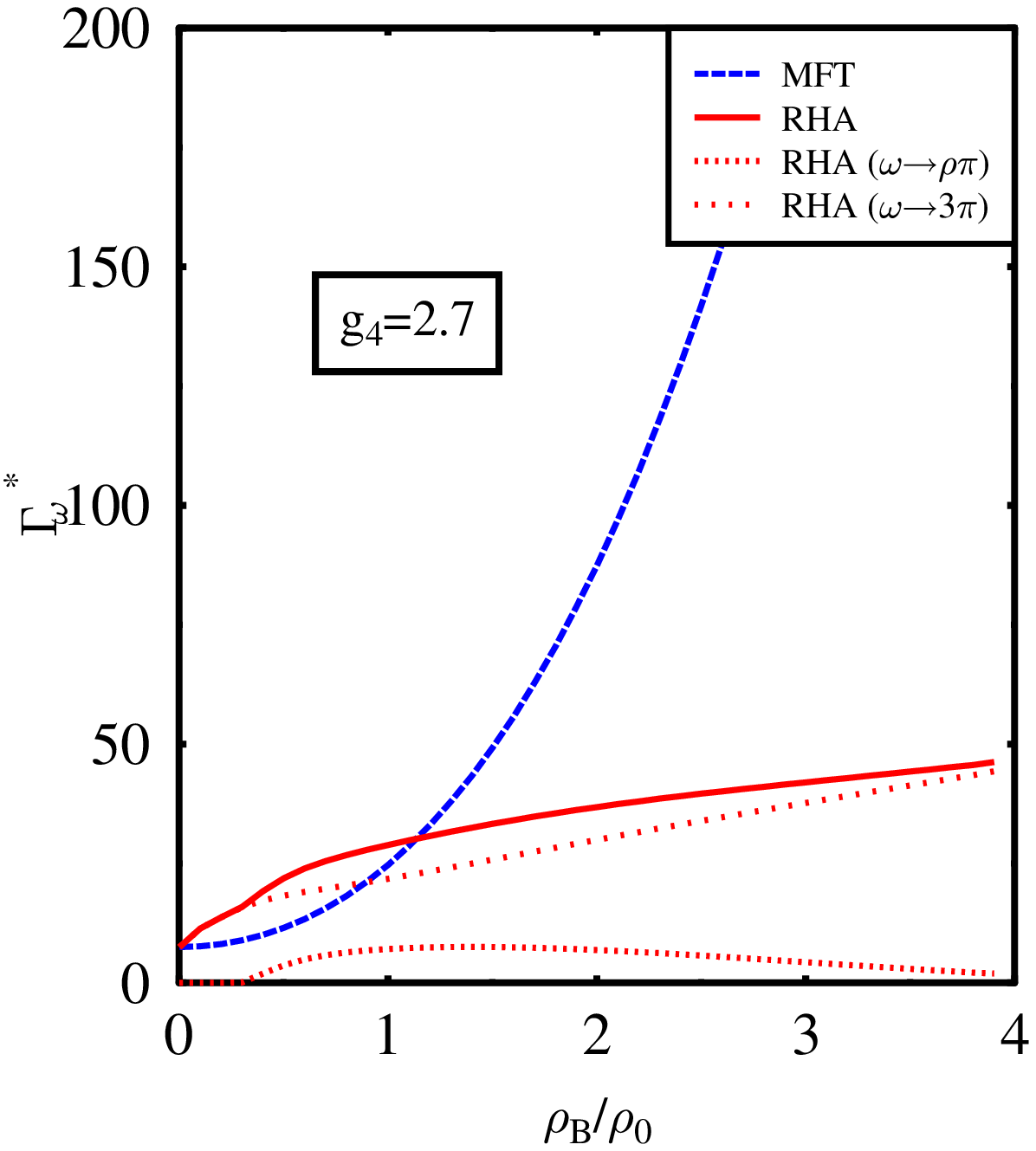}}}
\vspace{-1cm}
\caption{
\label{gmomgfig}
Effective decay width of $\omega$ meson without and with the
Hartree contributions. The decay width has contributions from
$\omega \rightarrow 3\pi$ as well as $\omega \rightarrow \rho \pi$.
The latter becomes accessible due to stronger medium modification of
the $\rho$ meson mass as compared to the $\omega$ mass. The MFT has no
contribution from the process $\omega \rightarrow \rho \pi$.
}
\end{center}
\end{figure}

The decay width of the $\omega$ meson is plotted
as a function of density in figure \ref{gmomgfig}. 
In the vacuum the process  
$\omega \rightarrow 3\pi$ is the dominant decay mode.
However, in the medium the channel  
$\omega \rightarrow \rho \pi$ also opens up, since the $\rho$-meson
has a stronger drop in the medium as compared to the $\omega$ meson mass.

The mean field approximation, does not have a contribution
from the latter decay channel, whereas the inclusion of relativistic
Hartree approximation permits both processes in the medium.
In the presence of the quartic vector meson interaction,
the channel
$\omega \rightarrow \rho \pi$, which opens up at around $0.3\rho_0$,
no longer remains kinematically accessible at higher densities. 
This is due to the 
increased importance of $\omega^4$ contributions to the vector meson
masses at high densities.   

The strong enhancement of the $\omega$ meson mass
in the presence of a quartic self interaction term for the $\omega$ field,
makes also the decay channel $\omega  \rightarrow N \bar N$ kinematically
accessible in the mean field approximation. 
In the present investigation
the vector meson properties are considered at rest. 
Vector mesons with a finite three momentum 
can also have additional decay channels to particle hole pairs.
These decay modes, e.g, have significant contributions, to the  
$\Delta$ decay width \cite{delta}.    
Additional channels that open up in the mean field approximation
in the presence of a quartic term for $\omega$, however, have not been
taken into consideration in the present work. 
Here the emphasis is on the effect due to the relativistic
Hartree approximation on the in-medium vector meson properties.

\begin{figure}
\begin{center}
\centerline{\parbox[b]{8cm}{
\includegraphics[width=9.2cm,height=9cm]{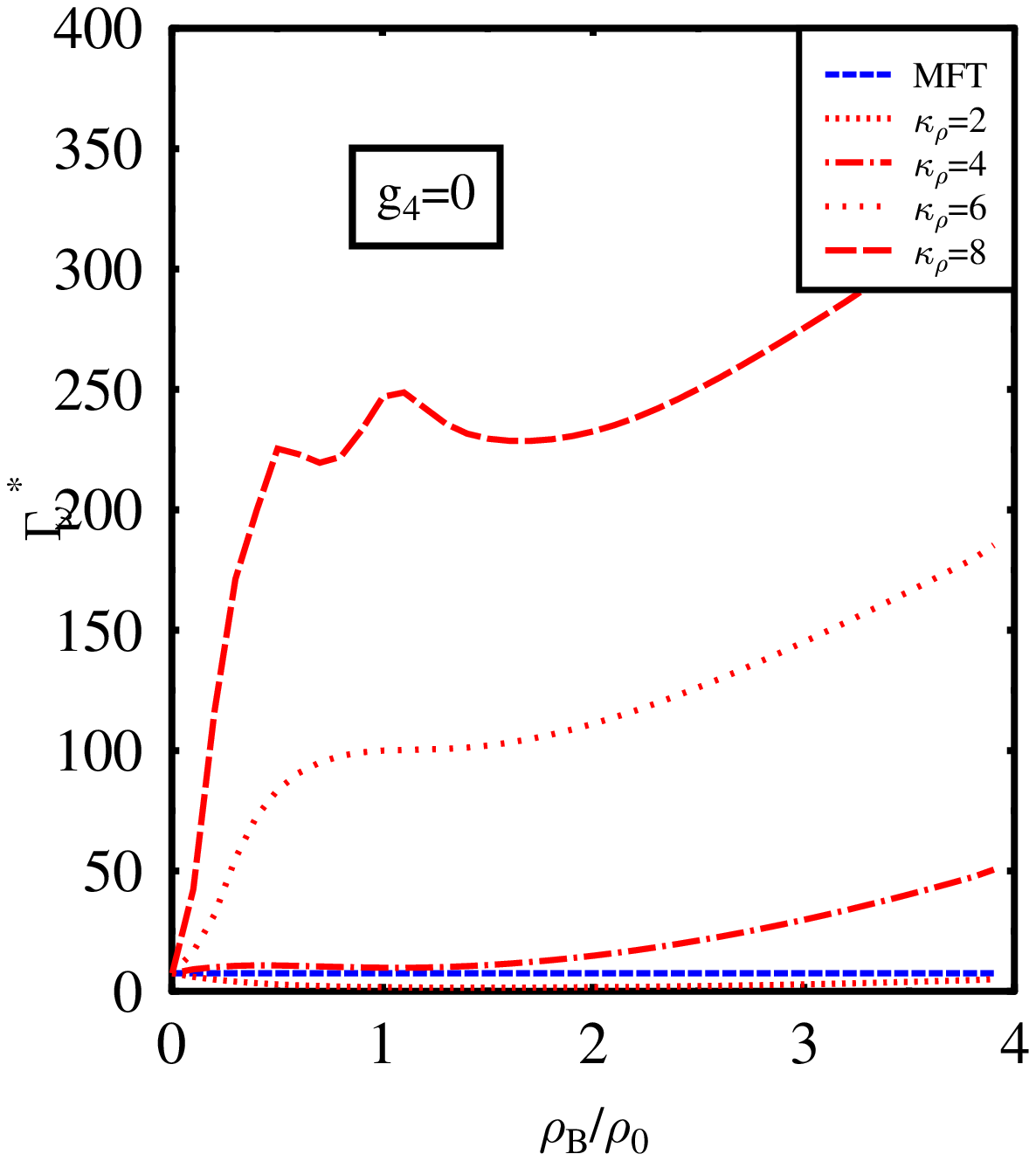}}
\parbox[b]{8cm}{
\includegraphics[width=9.2cm,height=9cm]{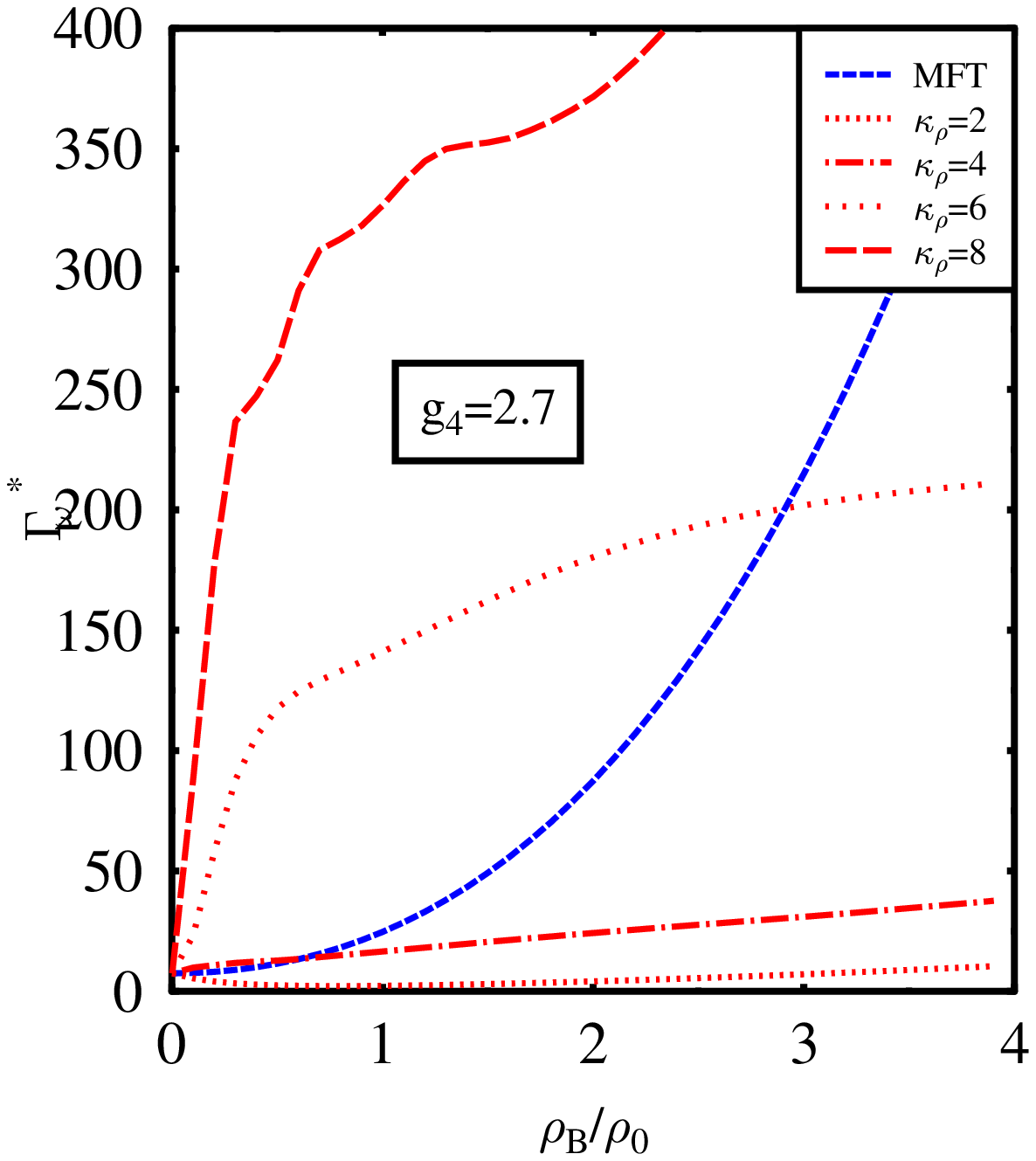}}}
\vspace{-1cm}
\caption{
\label{gmomgkfig}
Decay width of $\omega$ meson in the absence and presence of the
Dirac sea effect. For the channel $\omega \rightarrow \rho \pi$,
the $\rho$ meson properties are determined with the nucleon-rho vector
coupling, $g_{N\rho}$, as from the chiral model, which is compatible with
the symmetry energy and, the tensor coupling, $\kappa_\rho$ is taken
as a parameter.
}
\end{center}
\end{figure}

Figure \ref{gmomgkfig} illustrates the decay width of the $\omega$-meson
when the medium dependence of the $\rho N$ vector coupling is taken into
account and the tensor coupling is taken as a parameter. 
The strong dependence of the $\rho$
meson properties on the tensor coupling are reflected in the $\omega$-
decay width through the channel $\omega \rightarrow \rho \pi$.


\section{Summary}
\label{sum}
To summarize, in the present paper we have considered the modification
of the vector meson properties due to vacuum polarisation effects
arising from the Dirac sea in nuclear matter in the 
chiral SU(3) model. The baryonic properties as modified 
due to such effects determine the vector meson
masses in dense hadronic matter.
A significant reduction of these masses in the medium is found, 
where the Dirac sea contribution dominates over the Fermi sea part.
This shows the importance 
of the vacuum polarisation effects for the vector meson properties, 
as has been emphasized earlier 
within the framework of Quantum Hadrodynamics \cite{hatsuda1,jeans}. 

The $\rho$-meson mass is seen to have a sharper drop as compared to
the $\omega$-meson mass in the medium. This reflects the fact that the
vector meson-nucleon tensor coupling, which is absent for the
$\omega$-meson, plays an important role for the $\rho$-mass. The decay
width of $\rho \rightarrow \pi \pi$ is modified appreciably due to the
modification of the $\rho$ mass. 

The effects discussed above influence observables in
finite nuclei, stellar objects and 
relativistic heavy ion collisions.
For example the modified vector meson properties in a medium play an
important role in the dilepton emission rates 
in relativistic heavy ion collisions \cite{gale}. 
This is reflected by the shift and
broadening of the peaks in the low invariant mass regime in the
dilepton spectra. Therefore, it will be important to investigate 
how the dilepton rates are modified by 
the in-medium vector meson properties in the hot and dense hadronic matter.  
This necessitates the extension of the current work to finite
temperatures. Furthermore, it will be worthwhile to study, 
how the analysis of particle ratios for relativistic heavy ion 
collisions in \cite{zsch02} is affected by the Hartree contributions.
These and related problems are under investigation.

\begin{acknowledgements}
One of the authors (AM) is grateful to J. Reinhardt for fruitful
discussions and Institut fuer Theoretische Physik for warm
hospitality. This work is supported by Deutsche
Forschungsgemeinschaft (DFG), Gesellschaft f\"ur Schwerionenforschung
(GSI), Bundesministerium f\"ur Bildung und Forschung (BMBF), the
Graduiertenkolleg Theoretische und Experimentelle Schwerionenphysik
and by the U.S. Department of Energy, Nuclear Physics
Division (Contract No. W-31-109-Eng-38).
\end{acknowledgements}


\begin{thebibliography}{1}

\bibitem{helios} N. Masera for the HELIOS-3 collaboration,
Nucl. Phys. {\bf A 590}, 93c (1995).
\bibitem {ceres}G. Agakichiev et al (CERES collaboration),
Phys. Rev. Lett. {\bf 75}, 1272 (1995); 
G. Agakichiev et al (CERES collaboration), Phys. Lett. {\bf B 422},
405 (1998); G. Agakichiev et al (CERES collaboration), Nucl. Phys. 
{\bf A 661}, 23c (1999).
\bibitem {dls} R. J. Porter et al (DLS collaboration), Phys. Rev.
Lett. {\bf 79}, 1229 (1997); W. K. Wilson et al (DLS collaboration),
Phys. Rev. {C 57}, 1865 (1998).
\bibitem {rhic} D. P. Morrison (PHENIX collaboration), Nucl. Phys.
{\bf A 638}, 565c (1998).
\bibitem {hades} J. Stroth (HADES collaboration), Advances in 
Nuclear Dynamics {\bf 5}, 311 (1999). 
\bibitem {brown} G. E. Brown and M. Rho, Phys. Rev. Lett. {\bf 66}, 2720
(1991).
\bibitem{rapp}
R.~Rapp and J.~Wambach,
Adv.\ Nucl.\ Phys.\  {\bf 25} (2000) 1;
R.~Rapp, G.~Chanfray and J.~Wambach,
Nucl.\ Phys.\ A {\bf 617} (1997) 472.
\bibitem{hat} T. Hatsuda and Su H. Lee, Phys. Rev. {\bf C 46},
R34 (1992); T. Hatsuda, S. H. Lee and H. Shiomi, Phys. Rev. {\bf C 52},
3364 (1995).
\bibitem{jin} X. Jin and D. B. Leinweber, Phys. Rev. {\bf C 52}, 3344 
(1995); T. D. Cohen, R. D. Furnstahl, D. K. Griegel and X. Jin,
Prog. Part. Nucl. Phys. {\bf 35}, 221 (1995); R. Hofmann, Th. Gutsche,
A. Faessler, Eur. Phys. J. {\bf C 17}, 651 (2000); S. Mallik and
K. Mukherjee, Phys. Rev. {\bf D 58}, 096011 (1998).
\bibitem {samir} S. Mallik and A. Nyffeler, Phys. Rev. {\bf C 63},
065204 (2001).
\bibitem{weise} F. Klingl, N. Kaiser, W. Weise, Nucl. Phys. {\bf A 624},
527 (1997).
\bibitem{ernst} C. Ernst, S. A. Bass, M. Belkacem, H. St\"ocker
and W. Greiner, Phys. Rev. {\bf C 58}, 447 (1998).
\bibitem{qhd} B. D. Serot and J. D. Walecka, Adv. Nucl. Phys. {\bf 16},
1 (1986); S. A. Chin, Ann. Phys. (N. Y.) {\bf 108}, 301 (1977).
\bibitem{hatsuda} H. Shiomi and T. Hatsuda, Phys. Lett. {\bf B 334},
281 (1994).
\bibitem {hatsuda1} T. Hatsuda, H. Shiomi and H. Kuwabara, Prog. Theor.
Phys. {\bf 95}, 1009 (1996).
\bibitem{jeans} H.-C. Jean, J. Piekarewicz and A. G. Williams,
Phys. Rev. {\bf C 49}, 1981 (1994); K. Saito, K. Tsushima, A. W. Thomas,
A. G. Williams, Phys. Lett. {\bf B 433}, 243 (1998).
\bibitem{sourav} Jan-e Alam, S. Sarkar, P. Roy, B. Dutta-Roy and 
B. Sinha, Phys. Rev. {\bf C 59}, 905 (1999).
\bibitem{vecmass} A. Mishra, J. C. Parikh and W. Greiner, J. Phys. {\bf 
G 28}, 151 (2002).
\bibitem{dlp} A. Mishra, J. Reinhardt, H. St\"ocker and W. Greiner,
Phys. Rev. {\bf C 66}, 064902 (2002).
\bibitem{mishra} A. Mishra, P. K. Panda, S. Schramm, J. Reinhardt 
and W. Greiner, Phys. Rev. {\bf C 56}, 1380 (1997); 
A. Mishra, P. K. Panda and W. Greiner, J. Phys. {\bf G 27}, 1561 (2001);
A. Mishra, P. K. Panda and W. Greiner, J. Phys. {\bf G 28}, 67 (2002).

\bibitem{paper3}
P. Papazoglou, D. Zschiesche, S. Schramm, J. Schaffner-Bielich, H. St\"ocker,
  and W. Greiner, Phys. Rev. C {\bf 59},  411  (1999).

\bibitem{springer}
D. Zschiesche, P. Papazoglou, S. Schramm, C. Beckmann, J. Schaffner-Bielich, H.
  St\"ocker, and W. Greiner, Springer Tracts in Modern Physics {\bf 163},  129
  (2000).

\bibitem{saku69}
J.~J. Sakurai, {\em Currents and Mesons} (University of Chicago Press, Chicago,
  1969).

\bibitem{gasi69}
S. Gasiorowicz and D. Geffen, Rev. Mod. Phys. {\bf 41},  531  (1969).

\bibitem{mitt68}
P.~K. Mitter and L.~J. Swank, Nucl. Phys. B {\bf 8},  205  (1968).

\bibitem{toki}
Y. Sugahara and H. Toki, Nucl. Phys. {\bf A 579}, 557 (1994).

\bibitem{sche71}
J. Schechter and Y. Ueda, Phys. Rev. D {\bf 3},  168  (1971).

\bibitem{sche80}
J. Schechter, Phys. Rev. D {\bf 21},  3393  (1980).

\bibitem{serot97}
B.~D. Serot and J.~D. Walecka, Int. J. Mod. Phys. E {\bf 6},  515  (1997).

\bibitem{asakawa} M. Asakawa, C. M. Ko, P. Levai and X. J. Qiu,
Phys. Rev. {\bf C 46}, R1159 (1992); M. Herrmann, B. L. Friman
and W. N\"orenberg, Nucl. Phys. {\bf A 560}, 411 (1993);
G. Chanfray and P. Shuck, Nucl. Phys. {\bf A 545}, 271c (1992). 
\bibitem{grein} W. Grein, Nucl. Phys. {\bf B 131}, 255 (1977);
W. Grein and P. Kroll, Nucl. Phys. {\bf A 338}, 332 (1980).
\bibitem{sakurai} J. J. Sakurai, Currents and Mesons (The University
od Chicago Press, Chicago, 1969).
\bibitem{gellmann} M. Gell-Mann, D. Sharp, and W. D. Wagner, Phys.
Rev. Lett. {\bf 8}, 261 (1962). 
\bibitem{bali} B. A. Li, Phys. Rev. {\bf D 52}, 5165 (1995).
\bibitem{weisezp} F. Klingl, N. Kaiser and W. Weise, Z. Phys. {\bf A 356},
193 (1996).
\bibitem{kaymak} \"O. Kaymakcalan, S. Rajeev and I. Schechter, Phys.
Rev. {\bf D 30}, 594 (1984).

\bibitem {sym} N. K. Glendenning, F. Weber, S. A. Moszkowski,
Phys. Rev. {\bf C45}, 844 (1992).
\bibitem{delta} H. Kim, S. Schramm and S. H. Lee, Phys. Rev. {\bf C 56},
1582 (1997). 
\bibitem{gale} C. Gale, J. I. Kapusta, Nucl. Phys. {\bf B 357}, 65 (1991).
\bibitem{zsch02} D. Zschiesche, S. Schramm, J. Schaffner-Bielich,
H. St\"ocker, W. Greiner, Phys. Lett. {\bf B 547}, 7 (2002). 

\end{thebibliography}
\end{document}